\documentstyle[12pt,a41,epsfig,cite,amsmath]{article}
\newcommand\N{\nonumber}

\newcommand{\bea}{\begin{eqnarray}}
\newcommand{\bq}{\begin{equation}}
\newcommand{\eea}{\end{eqnarray}}
\newcommand{\eq}{\end{equation}}

\newcommand{\gsim}{\raisebox{-0.07cm   }
{$\, \stackrel{>}{{\scriptstyle\sim}}\, $}}
\newcommand\ds{\displaystyle}

\newcommand\GeV{\,\mbox{GeV}}

\newcommand\be{\begin{eqnarray}}
\newcommand\ee{\end{eqnarray}}

\newcommand\ep{\varepsilon}

\newcommand\Li{{\rm Li}}

\newcommand\Mvec{\,\mbox{\bf M}}
\begin{document}
\renewcommand{\thefootnote}{\fnsymbol{footnote}}
\noindent
\sloppy
\thispagestyle{empty}
\begin{flushleft}
DESY 07-026 \hfill
{\tt hep-ph/0703285}\\
SFB/CPP--07--11\\
March 2007
\end{flushleft}
%
\vspace*{\fill}
\begin{center}
{\LARGE\bf Two--Loop Massive Operator Matrix Elements}

\vspace{2mm}
{\LARGE\bf and Unpolarized Heavy Flavor 
Production at}

\vspace{2mm}
{\LARGE\bf Asymptotic Values \boldmath{$Q^2 \gg m^2$}~\footnote{\sf 
Dedicated to the Memory of W.L. van Neerven}}

\vspace{2cm}
\large
Isabella Bierenbaum, Johannes Bl\"umlein  and
Sebastian Klein
\\
\vspace{2em}
\normalsize
{\it Deutsches Elektronen--Synchrotron, DESY,\\
Platanenallee 6, D--15738 Zeuthen, Germany}
\\
\vspace{2em}
\end{center}
\vspace*{\fill}
%
\begin{abstract}
\noindent
We calculate the $O(\alpha_s^2)$ massive operator matrix elements for the
twist--2 operators, which contribute to the heavy flavor Wilson coefficients
in unpolarized deeply inelastic scattering in the region $Q^2 \gg m^2$. 
The calculation has been performed using light--cone  expansion techniques.
We confirm an earlier result obtained in \cite{Buza:1995ie}. The calculation 
is carried out without using the integration-by-parts method and in Mellin 
space using harmonic sums, which lead to a significant compactification of 
the analytic results derived previously. The results allow to determine the 
heavy flavor Wilson coefficients for $F_2(x,Q^2)$ to $O(\alpha_s^2)$ and 
for $F_L(x,Q^2)$ to $O(\alpha_s^3)$ for all but the power suppressed terms 
$\propto (m^2/Q^2)^k, k \geq 1$. 
\end{abstract}
\vspace*{\fill}
\newpage
\section{Introduction}
\label{sec:introduction}

\vspace{1mm}\noindent
Deeply inelastic electron--nucleon scattering at large momentum transfer
allows to measure the parton distribution functions of the nucleons together 
with the QCD scale $\Lambda_{\rm QCD}$. In the region of large hadronic masses
$W^2 \simeq Q^2 (1-x)/x$ the sea--quark distribution receives substantial
contributions due to heavy flavor (charm and beauty) pair production.
At the level of leading twist $\tau = 2$ their contribution to the deeply 
inelastic structure functions is described by heavy quark Wilson 
coefficients, which are convoluted with the light quark and gluon parton 
densities. Depending on the range of the Bjorken variable $x$ and the 
gauge boson virtuality $Q^2$, these contributions can amount to 20--40\% 
of the structure functions~\cite{STRAT}. The unpolarized heavy flavor 
Wilson coefficients were calculated at leading order (LO) in Refs.~\cite{LO}. The 
next-to-leading order (NLO) corrections were derived in semi-analytic form in 
Refs.~\cite{NLO} in $x$--space for the structure functions $F_2(x,Q^2)$ and $F_L(x,Q^2)$. 
A fast numerical implementation in Mellin space was given in 
\cite{Alekhin:2003ev}. For the asymptotic region $Q^2 \gg m^2$, an analytic 
result for the heavy flavor coefficient functions was calculated to $O(\alpha_s^2)$ 
in \cite{Buza:1995ie}. In the case of the structure function $F_L(x,Q^2)$, 
the asymptotic result to $O(\alpha_s^3)$ was derived in \cite{Blumlein:2006mh}. 
The leading order small-$x$ resummation for 
$F_{2,L}^{Q\overline{Q}}(x,Q^2)$ was calculated in~\cite{Catani:1990eg}.
The heavy quark Wilson coefficients differ significantly from those 
of the light quarks even in the asymptotic region. Therefore, the scaling 
violations of the heavy flavor part in $F_{2,L}(x,Q^2)$ are different 
from those of the light flavor contributions. Both for the measurement of the QCD scale 
$\Lambda_{\rm QCD}$ and for the extraction of the light parton densities a correct 
description of the heavy flavor contributions is therefore required.
As shown in Ref.~\cite{Buza:1995ie}, in case of the structure function $F_2(x,Q^2)$ the asymptotic heavy 
flavor 
terms describe the complete contributions very well already for scales $Q^2 \gsim 30 \GeV^2$, whereas 
for $F_L(x,Q^2)$ this applies only at much higher scales, $Q^2 \gsim 800 
\GeV^2$.

In the present paper, we recalculate for the first time the asymptotic 2--loop 
corrections to the heavy flavor structure functions using a different 
method than in Ref.~\cite{Buza:1995ie}. 
All logarithmic terms and the constant term of the heavy flavor Wilson coefficients are 
obtained due to a factorization of this quantity into the massive 
operator matrix elements and the light parton Wilson coefficients, which are known 
from the literature \cite{LWILS1,LWILS2,LWILS3}. 
In \cite{Buza:1995ie} the massive operator matrix elements were derived in 
momentum-fraction $(x)$-space. The corresponding 2--loop integrals were 
simplified using the integration-by-parts method \cite{Chetyrkin:1980pr}.
We will work in Mellin-space, accounting for the appropriate symmetry of 
the problem, and do thoroughly avoid the integration-by-parts method.
This requires to solve more complicated Feynman-parameter integrals. However,
in this way we are able to derive by far more compact results, even for the 
individual Feynman diagrams. In the direct calculation, we use 
Mellin--Barnes integrals \cite{MB1,MB2} and representations through 
generalized hypergeometric functions \cite{HGF}. A brief account on scalar 
2--loop integrals to be derived in the present calculation was given in
\cite{Bierenbaum:2006mq,Bierenbaum:2007dm,SKLEIN}. The final expressions obtained
allow to represent the asymptotic heavy flavor contributions 
to the deep--inelastic structure functions in Mellin space in a completely 
analytic form. Precise representations of the analytic continuations of
the harmonic sums w.r.t. the Mellin index $N$ to complex variables are given
in \cite{ANCONT1,JB07}. A fast single numerical inverse Mellin transformation yields
the structure functions in $x$-space. This representation is well suited
for fast numerical data analysis \cite{EVOL}. 

The paper is organized as follows. In Section~2, we describe the 
principal method applied to derive the 2--loop corrections in the 
asymptotic region $Q^2 \gg m^2$, covering all contributions but the power 
corrections $\propto (m^2/Q^2)^k,~~k \geq 1$. In Section~3, the renormalization
of the massive operator matrix elements is described.
Section~4 gives a brief account of the 1--loop corrections. The 2--loop corrections 
to the operator matrix elements are derived in Section~5. 
Working in $D = 4 + \ep$ space--time dimensions, the splitting functions,
related to the problem, can be unfolded in leading and next-to-leading order,
which provides a check for the calculation. Here we also discuss the mathematical 
structure of the results and compare to the result obtained in Ref.~\cite{Buza:1995ie}. 
Section~6 contains the conclusions. The appendices summarize details of the calculation 
and different types of summation formulae used in the present calculation,   which are of general 
interest for other higher order calculations. 
\section{\boldmath The Method}
\label{sec:METH}

\vspace{1mm}\noindent
In the twist--2 approximation, the deep--inelastic nucleon structure functions 
$F_i(x,Q^2),~i=2,L,$
are described as Mellin convolutions between the parton densities 
$f_j(x,\mu^2)$ and the Wilson coefficients ${\mathsf{C}}_i^j(x,Q^2/\mu^2)$
\begin{eqnarray}
\label{STR}
F_i(x,Q^2) = \sum_j {\mathsf{C}}_i^j\left(x,\frac{Q^2}{\mu^2}\right) \otimes 
f_j(x,\mu^2)
\end{eqnarray}
to all orders in perturbation theory due to the factorization theorem. 
Here $\mu^2$ denotes the factorization scale and
the Mellin convolution is given by the integral
\begin{eqnarray}
[A \otimes B](x) = \int_0^1 dx_1 \int_0^1 dx_2~~ \delta(x - x_1 x_2) 
~A(x_1) B(x_2)~.
\end{eqnarray}
Since the distributions $f_j$ refer to {\sf massless} partons, the heavy 
flavor effects are contained in the Wilson coefficients only.
We will derive the massive contributions in the region $Q^2 \gg 
m^2$. These are the non--power corrections in $m^2/Q^2$, i.e. all 
logarithmic contributions and the constant term. We apply the collinear 
parton model, i.e. the parton 4--momentum is $p = x P$, with  
$P$ the nucleon momentum. The massive Wilson coefficients itself
can be viewed as a quasi cross section in $p V^*$ scattering, where $V^*$ 
denotes the exchanged virtual vector boson. In the limit $Q^2 \gg m^2$, the 
massive Wilson coefficients $H^{\rm S,NS}_{2,L,i}(Q^2/m^2,m^2/\mu^2,x)$
factorize \cite{Buza:1995ie}
into Wilson coefficients $C_{2,L;k}^{S,NS} \left(Q^2/\mu^2,x\right)$ 
accounting for light flavors only and  massive operator matrix 
elements $A_{k,i}^{\rm S, NS} \left(m^2/\mu^2,x\right)$. 
\begin{eqnarray}
\label{HFAC}
H_{2,L;i}^{\rm S, NS}\left(\frac{Q^2}{m^2}, \frac{m^2}{\mu^2},x\right) = 
C_{2,L;k}^{S,NS} \left(\frac{Q^2}{\mu^2},x\right) 
\otimes	A_{k,i}^{\rm S, NS} \left(\frac{m^2}{\mu^2},x\right) 
\end{eqnarray}
The 
latter take a similar role as the parton densities in (\ref{STR}). They are process independent 
but perturbatively calculable. The factorization (\ref{HFAC}) is a 
consequence of the renormalization group equation.
The operator matrix elements $A_{k,i}^{\rm S, NS}$ obey the expansion
\begin{eqnarray}
\label{op1}
A_{k,i}^{\rm S, NS} \left(\frac{m^2}{\mu^2}\right) = \langle 
i|O_k|i\rangle
= \delta_{k,i} + \sum_{l=1}^{\infty} a_s^l A_{k,i}^{{\rm S, 
NS},(l)},~~~~i=q,g 
\end{eqnarray}
of the twist--2 quark singlet and non--singlet  operators $O_k^{\rm S, 
NS}$ between {\sf partonic} states 
$|i\rangle$, 
which are related by collinear factorization to the initial--state nucleon 
states $|N\rangle$. $a_s = \alpha_s(\mu^2)/(4\pi)$ denotes the strong 
coupling constant. The Feynman rules for the operator insertions are given in
Figure~1.
Since the operator matrix elements are process--independent 
quantities, the process dependence of $H_{2,L;i}^{\rm S, NS}$ 
is described by the associated light parton coefficient functions 
\begin{eqnarray}
\label{co1}
{C}_{2,L;k} \left(\frac{Q^2}{\mu^2}\right) = 
\sum_{l=l_0}^{\infty} a_s^l C_{L,k}^{(l)}
\left(\frac{Q^2}{\mu^2}\right), \hspace{7mm} k= {\rm NS, S}, g~.
\end{eqnarray}
The $\overline{\rm MS}$ 
coefficient (and splitting) functions, in the massless limit, corresponding to the heavy 
quarks only, are denoted by
\begin{eqnarray}
\widehat{C}_{2,L;k}\left(\frac{Q^2}{\mu^2}\right) = 
C_{2,L;k}\left(\frac{Q^2}{\mu^2},N_L+N_H\right)  
- C_{2,L;k}\left(\frac{Q^2}{\mu^2},N_L\right)~, 
\end{eqnarray}
where $N_H, N_L$ are the number of heavy and light flavors, 
respectively. In the following we will consider the case of a single heavy 
quark, i.e. $N_H = 1$. The formalism is easily generalized to more than 
one heavy quark species.

The massive operator matrix elements to $O(a_s^2)$ allow to calculate 
the heavy quark Wilson coefficients in the asymptotic region for $F_2(x,Q^2)$ 
to $O(a_s^2)$ \cite{Buza:1995ie} and for $F_L(x,Q^2)$ to $O(a_s^3)$ \cite{Blumlein:2006mh}.
The general structure of the Wilson coefficients is
\begin{eqnarray}
\label{eqH2g}
H_{2,g}^{\rm S}\left(\frac{Q^2}{m^2}, \frac{m^2}{\mu^2}\right)
&=& a_s    \left[\widehat{C}^{(1)}_{2,g}\left(\frac{Q^2}{\mu^2}\right) 
               + A_{Qg}^{(1)}\left(\frac{\mu^2}{m^2}\right)\right] \nonumber\\
&+&   a_s^2  \left[ 
                  \widehat{C}^{(2)}_{2,g}\left(\frac{Q^2}{\mu^2}\right)
+A_{Qg}^{(1)}\left(\frac{\mu^2}{m^2}\right) \otimes
                  C^{(1)}_{2,q}\left(\frac{Q^2}{\mu^2}\right)
+A_{Qg}^{(2)}\left(\frac{\mu^2}{m^2}\right) 
\right]
\\
H_{2,q}^{\rm PS}\left(\frac{Q^2}{m^2}, \frac{m^2}{\mu^2}\right)
&=& a_s^2  \left[\widehat{C}^{{\rm PS},(2)}_{2,q}\left(\frac{Q^2}{\mu^2}\right)
 +              A_{Qq}^{{\rm PS},(2)}\left(\frac{\mu^2}{m^2}\right) \right]
\\
H_{2,q}^{\rm NS}\left(\frac{Q^2}{m^2}, \frac{m^2}{\mu^2}\right)
&=& a_s^2  \left[\widehat{C}^{{\rm NS},(2)}_{2,q,Q}\left(\frac{Q^2}{\mu^2}\right)
 +              A_{qq,Q}^{{\rm NS},(2)}\left(\frac{\mu^2}{m^2}\right) \right]
\\
\label{eq1a}
H_{L,g}^{\rm S}\left(\frac{Q^2}{m^2}, \frac{m^2}{\mu^2}\right)
&=& a_s    \widehat{C}^{(1)}_{L,g}\left(\frac{Q^2}{\mu^2}\right)
 +  a_s^2  \left[ 
A_{Qg}^{(1)}\left(\frac{\mu^2}{m^2}\right) \otimes
                  C^{(1)}_{L,q}\left(\frac{Q^2}{\mu^2}\right)
 +                \widehat{C}^{(2)}_{L,g}\left(\frac{Q^2}{\mu^2}\right)\right]
\nonumber\\
&+& a_s^3 \left[
A_{Qg}^{(2)}\left(\frac{\mu^2}{m^2}\right) \otimes
                  C^{(1)}_{L,q}\left(\frac{Q^2}{\mu^2}\right)
+ A_{Qg}^{(1)}\left(\frac{\mu^2}{m^2}\right) \otimes
                  C^{(2)}_{L,q}\left(\frac{Q^2}{\mu^2}\right)
 +
\widehat{C}^{(3)}_{L,g}\left(\frac{Q^2}{\mu^2}\right)\right]
\nonumber\\ \\
\label{eq1b}
H_{L,q}^{\rm PS}\left(\frac{Q^2}{m^2}, \frac{m^2}{\mu^2}\right)
&=& a_s^2  \widehat{C}^{{\rm PS},(2)}_{L,q}\left(\frac{Q^2}{\mu^2}\right)
 +  a_s^3  \left[
A_{Qq}^{{\rm PS},(2)}\left(\frac{\mu^2}{m^2}\right) \otimes
                  C^{(1)}_{L,q}\left(\frac{Q^2}{\mu^2}\right)
 +\widehat{C}^{{\rm PS},(3)}_{L,q}\left(\frac{Q^2}{\mu^2}\right)\right]
\\
\label{eqHLqNS}
H_{L,q}^{\rm NS}\left(\frac{Q^2}{m^2}, \frac{m^2}{\mu^2}\right)
&=& a_s^2    \widehat{C}^{{\rm NS},(2)}_{L,q}\left(\frac{Q^2}{\mu^2}\right)
 +  a_s^3  \left[
A_{qq,Q}^{{\rm NS},(2)}\left(\frac{\mu^2}{m^2}\right) \otimes
                  C^{(1)}_{L,q}\left(\frac{Q^2}{\mu^2}\right)
 +
\widehat{C}^{{\rm NS},(3)}_{L,q}\left(\frac{Q^2}{\mu^2}\right)\right].
\end{eqnarray}
%
\section{\bf\boldmath Renormalization}
\label{Sec-RegRen}
%
The calculation is performed in $D= 4 + \varepsilon$ 
dimensions in the $\overline{\rm MS}$--scheme. 
The massive operator matrix elements $A_{ij}(m^2/\mu^2)$, with 
$\mu$ being the renormalization scale, exhibit ultraviolet divergences which have to be 
removed by operator--, mass-- and coupling constant renormalization. 
Furthermore, it contains collinear singularities, in the present case 
beginning with 2--loop order. 
We follow the notation of Ref.~\cite{Buza:1995ie} and briefly summarize 
the renormalization procedure. 

The external lines of the diagrams are treated on--shell after 
factorization. The scale for the process is set by the heavy quark mass 
$m$. The $Z_{O,ik}$--matrix performs the 
renormalization of the composite 
operator turning $\hat{A}_{ij}(m^2/\mu^2,{a}_s,\varepsilon)$
into $\tilde{A}_{ij}(m^2/\mu^2,{a}_s,\varepsilon)$,
\begin{eqnarray}
\hat{A}_{ij}(m^2/\mu^2,{a}_s,\varepsilon)
= Z_{O,jk}(a_s, \varepsilon) \otimes
\tilde{A}_{kj}(m^2/\mu^2,{a}_s,\varepsilon)~,
\end{eqnarray}
with
\begin{eqnarray}
Z_{O,ij}      &=& \sum_{k=0}^\infty a_s^k  Z_{O,ij}^k,\\
\gamma_{ij}^N &=& - (Z_{O}(\mu))^{-1}_{ik} \frac{\partial}{\partial \mu}
Z_{O,kj}(\mu)~.
\end{eqnarray}
Here, $\gamma_{ij}^N$ denote the singlet anomalous dimensions, which are related to
the splitting functions by
\begin{eqnarray}
\label{eqAN}
\gamma_{ij}^N = - \int_0^1 dz z^{N-1} P_{ij}(z)~.
\end{eqnarray}
The collinear singularities are removed by the matrices $\Gamma_{kj}(a_s,\varepsilon)$,
\begin{eqnarray}
\Tilde{A}_{ij}(m^2/\mu^2,{a}_s,\varepsilon)
=\Tilde{\Tilde{A}}_{ik}(m^2/\mu^2,{a}_s,\varepsilon)
\otimes \Gamma_{kj}(a_s, \varepsilon) 
\end{eqnarray}
with
\begin{eqnarray}
\Gamma_{ij}      &=& \sum_{k=0}^\infty a_s^k  \Gamma_{ij}^k~.
\end{eqnarray}
To 2--loop order, one has
\begin{eqnarray}
\label{eqGAM}
\Gamma_{ij}      &=& \delta_{ij} + a_s S_\varepsilon 
\frac{1}{\varepsilon} P_{ij}^{(0)} \nonumber\\ & &
+ a_s^2 S_\varepsilon^2\left[\frac{1}{\varepsilon^2} \left\{\frac{1}{2} 
P_{ik}^{(0)} \otimes P_{kj}^{(0)} + \beta_0 P_{ij}^{(0)} \right\} + 
\frac{1}{2 \varepsilon} P_{ij}^{(1)}\right] + O(a_s^3)~.
\end{eqnarray}
The factorization and renormalization scales are choosen to be equal, $\mu_R = \mu_F \equiv \mu$.
The spherical factor 
$S_\varepsilon$ collects all terms to be removed in the $\overline{\rm 
MS}$ scheme
\begin{eqnarray}
S_\varepsilon = \exp \left\{\frac{\varepsilon}{2} [\gamma_E - \ln(4\pi)] 
\right\}~.
\end{eqnarray}
$\gamma_E$ is the Euler--Mascheroni constant. 

Finally, the mass and coupling constant renormalization has to be carried 
out.
The bare coupling $\hat{a}_s$ and bare mass $\hat{m}$ are related to the 
renormalized quantities by
\begin{eqnarray}
\hat{a}_s &=& Z_g^2 a_s  = a_s(\mu^2) \left[1 + a_s(\mu^2) \cdot \delta 
a_s\right] + O(a_s^3)~,\\
\hat{m}   &=& Z_m m = a_s \delta m + O(a_s^2)~.
\end{eqnarray}
We choose the on--mass--shell scheme for mass renormalization. Here,
\begin{eqnarray}
\label{eqZG}
Z_g &=& \frac{Z_1^l + Z_1^H}{(Z_3^l+Z_3^H)^{3/2}}
= 1 + a_s S_\varepsilon \left\{ \frac{\beta_0}{\varepsilon}
+ \frac{\beta_{0,Q}}{\varepsilon} \left[1 +\frac{\zeta_2}{8} \varepsilon^2\right] \sum_{N_H=4}^6 
\left(\frac{m_{N_H}^2}{\mu^2}\right)^{\varepsilon/2} \right\}~,\\
\label{eqZM}
Z_m &=&  1 + \hat{a}_s C_F S_\varepsilon \left(\frac{m^2}{\mu^2}\right) ^{\varepsilon/2}
\left[\frac{6}{\varepsilon} - 4 \right]~,
\end{eqnarray}
in Feynman gauge \cite{Gross:1973id,Politzer:1973fx,Gastmans:1973uv}. $N_H$ denotes the heavy quark 
species and the $SU(3)_c$ color factors are 
$C_F = (N_c^2-1)/(2 N_c), C_A = N_c, T_R = 1/2, N_c = 3$. The $Z$--factors in (\ref{eqZG}, \ref{eqZM})
read~:
\begin{eqnarray}
Z_1^l &=& 1 + a_s\frac{2}{\varepsilon} S_\varepsilon \left[ -\frac{2}{3} C_A + \frac{4}{3} T_R N_f \right]\\
Z_3^l &=& 1 + a_s\frac{2}{\varepsilon} S_\varepsilon \left[ -\frac{5}{3} C_A + \frac{4}{3} T_R N_f \right]\\
Z_1^H &=& Z_3^H = a_s \frac{S_\varepsilon}{\varepsilon} \frac{8}{3} T_R \left[1 + \frac{\zeta_2}{8} 
\varepsilon^2\right] \sum_{N_H=4}^6 \left(\frac{m_{N_H}^2}{\mu^2}\right)^{\varepsilon/2}~.
\end{eqnarray}
The lowest expansion coefficients of the $\beta$-functions are
\begin{eqnarray}
\beta_0     &=& \frac{1}{3} \left[ 11 C_A- 4 T_R N_f\right],
\\
\beta_{0,Q} &=& - \frac{4}{3} T_R~.
\end{eqnarray}
The bare and renormalized coupling and quark mass are related by
\begin{eqnarray}
\delta a_s &=& S_\varepsilon \left[ \frac{2 \beta_0}{\varepsilon}
+ \sum_{N_H=4}^6 \frac{2 \beta_{0,Q}}{\varepsilon} 
\left(\frac{m^2_{N_H}}{\mu^2}\right)^{\varepsilon/2} \left( 1 + \frac{1}{8} 
\varepsilon^2 \zeta_2\right)\right]~,\\
\delta m &=& C_F S_\varepsilon m \left(\frac{m^2}{\mu^2} 
\right)^{\varepsilon/2} \left\{\frac{6}{\varepsilon} -4\right\}~.
\end{eqnarray}
The renormalized operator matrix element $A_{ij}$ 
is now given by
\begin{eqnarray}
\label{eqOM}
A_{ij} &=& \delta_{ij} 
           + a_s \left[
\hat{A}_{ij}^{(1)} + Z_{O,ij}^{-1,(1)}
+ \Gamma_{ij}^{-1,(1)} 
                 \right] \nonumber\\ & &
+ a_s^2 \Bigl[ 
  \hat{A}_{ij}^{(2)} + \delta m \frac{d}{dm} A_{ij}^{(1)} +
  \delta a_s \hat{A}_{ij}^{(1)} + Z_{O,ik}^{-1,(1)} \otimes \hat{A}_{kj}^{(1)}
+Z_{O,ij}^{-1,(2)}  \nonumber\\ & &
 + \left\{\hat{A}_{ik}^{(1)} + Z_{O,ik}^{-1,(1)} \right\} \otimes 
\Gamma^{-1,(1)}_{kj}
+\Gamma^{-1,(2)}_{kj} \Bigr] + O(a_s^3)~.
\end{eqnarray}

In the following sections we first calculate the un-renormalized operator 
matrix elements $\hat{A}_{ij}$ from which $A_{ij}$ is derived through 
(\ref{eqOM}). Due to (\ref{eqAN}, \ref{eqGAM}) the following leading and 
next-to-leading order splitting functions are needed. We will 
mainly work 
in Mellin space and therefore list these functions in this 
representation. The leading order splitting functions read \cite{LOSP}
\begin{eqnarray}
\label{eqPqq0}
P_{qq}^{(0)}(N) &=& 4 C_F\left[-2S_1(N-1) + \frac{(N-1) (3N+2)}{2 N
(N+1)}\right]\\
\label{eqPqg0}
P_{qg}^{(0)}(N) &=& 8 T_R N_F \frac{N^2 + N + 2}{N (N+1) (N+2)}
\\
\label{eqPgg0}
P_{gg}^{(0)}(N) &=& 8 C_A\left[-S_1(N-1) - \frac{N^3 - 3 N -4}
{(N-1) N (N+1) (N+2)}\right] + 2 \beta_0
\\
\label{eqPgq0}
P_{gq}^{(0)}(N) &=& 4 C_F \frac{N^2 + N +2}{(N-1) N (N+1)}
\end{eqnarray}
Furthermore, the following next-to-leading order splitting functions
contribute \cite{NLOSP,FLO}
\begin{eqnarray}
\label{eqPqqPS}
\widehat{P}_{qq}^{{\rm PS},(1)}(N) &=& 16 C_F T_R  \frac{5 N^5 + 32 N^4
                           + 49 N^3 +38 N^2 +28 N + 8}{(N-1) N^3 (N+1)^3
(N+2)^2}
\\
\label{eqPqqQ}
P_{qq,Q}^{{\rm NS},(1)}(N) = \widehat{P}_{qq}^{{\rm NS},(1)} &=& C_F T_R
\left\{\frac{160}{9} S_1(N-1) -
\frac{32}{3} S_2(N-1)\right.\nonumber\\
&& \hspace{12mm} \left. -\frac{4}{9}\frac{(N-1)(3N+2)(N^2-11 N -6)}{N^2
(N+1)^2}\right\}\\
\label{eqPqg1}
\widehat{P}_{qg}^{(1)}(N) &=& 8 C_F T_R \left\{
                    2 \frac{N^2 + N + 2}{N (N+1) (N+2)} \left[S_1^2(N)
                    - S_2(N)\right] - \frac{4}{N^2} S_1(N)
\right. \nonumber
\end{eqnarray} \begin{eqnarray}
& & \left. \hspace{1.3cm} + \frac{5 N^6 + 15 N^5 + 36 N^4 + 51 N^3 +25 N^2
+ 8 N +4}{N^3 (N+1)^3 (N+2)}\right\}  \nonumber\\
& &+  16 C_A T_R \left\{
-\frac{N^2+N+2}{N(N+1)(N+2)} \left[S_1^2(N)
+ S_2(N) - \zeta_2 - 2 \beta'(N+1) \right]
\right. \nonumber\\
& & \hspace{1.3cm} \left. + 4 \frac{2N +3}{(N+1)^2 (N+2)^2} S_1(N)
 + \frac{P_1(N)}{(N-1) N^3 (N+1)^3 (N+2)^3}\right\}~,
\nonumber\\
\end{eqnarray}
where
\begin{eqnarray}
P_1(N) &=& N^9 +6 N^8 + 15 N^7 + 25 N^6 + 36 N^5 + 85 N^4 + 128 N^3
+ 104 N^2  \nonumber\\ & &
+ 64 N +16~.
\end{eqnarray}
Here, the harmonic sums~\cite{HSUM1,HSUM2} are given by
\begin{eqnarray}
S_1(N)         &=& \sum_{l=1}^N \frac{1}{l} = \psi(N+1) +\gamma_E 
\\
S_k(N)         &=& \sum_{l=1}^N \frac{1}{l^k} = \frac{(-1)^{k-1}}{(k-1)!} \psi^{(k-1)}(N+1) +\zeta_k,~~~k \geq 2 
\\
\beta(N) &=& \frac{1}{2} \left[\psi\left(\frac{N+1}{2}\right) - \psi\left(\frac{N}{2}\right)\right] \\
S_{-1}(N) &=& (-1)^{N} \beta(N+1) - \ln(2)
\\
S_{-k}(N) &=& \frac{(-1)^{N+k+1}}{(k-1)!} \beta^{k-1}(N+1) - 
\left(1-\frac{1}{2^{k-1}}\right) \zeta_k,~~~k \geq 2~.
\end{eqnarray}
$\zeta_k$ denotes the Riemann $\zeta$--function.
 \section{\bf\boldmath 
The one-loop massive operator matrix elements}
          \label{Sec-OPE1loop}
\begin{figure}[htb]
\begin{center}
\includegraphics[angle=0, width=14.0cm]{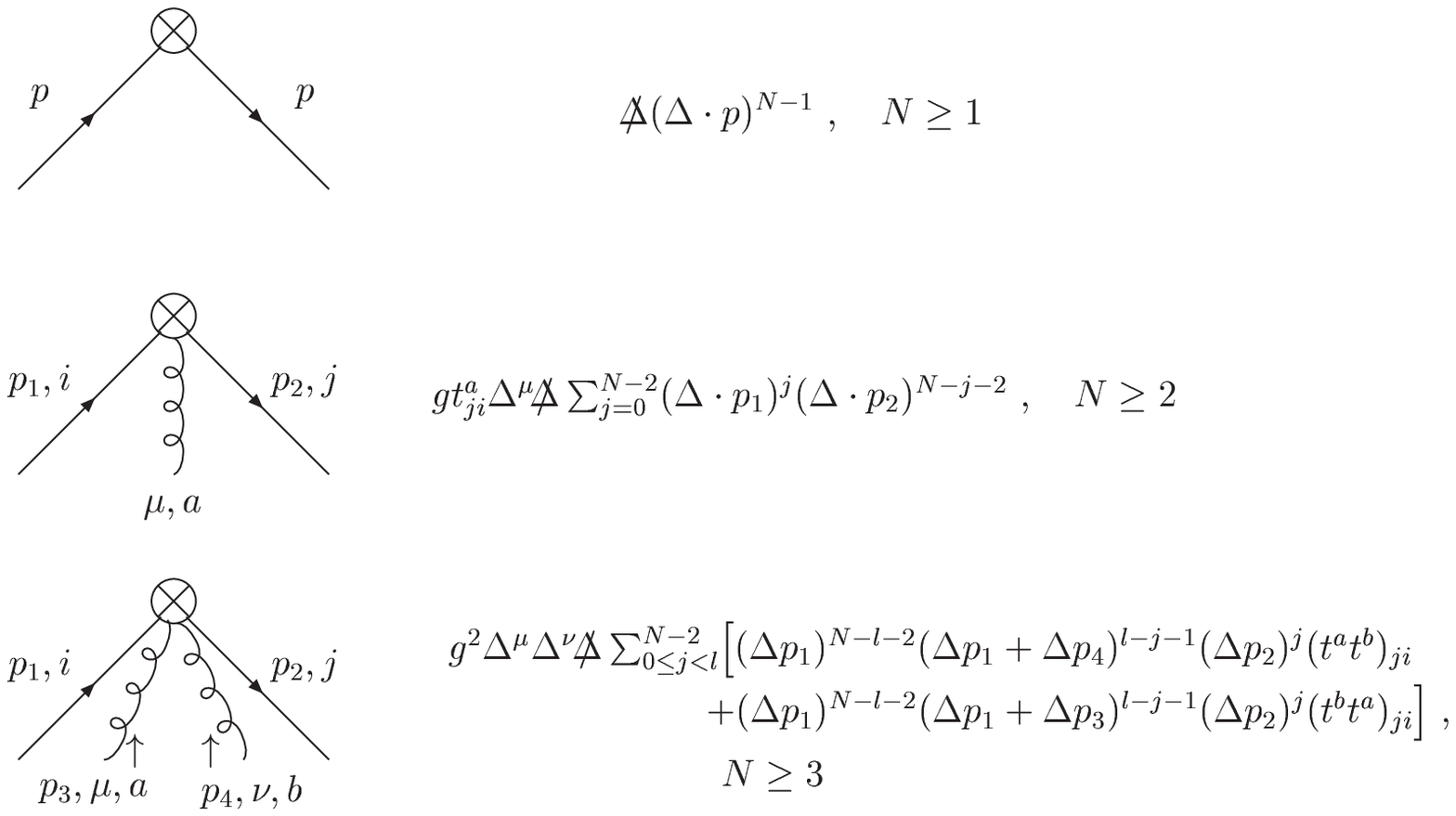} 
\end{center}
\caption{\label{fig:1}
\sf 
Feynman rules for the operator insertion $\otimes$ to $O(a_s^2)$, cf.~\cite{FLO}.
$\Delta$ denotes a light--like vector, $\Delta.\Delta = 0$.
}
\end{figure}

\begin{figure}[htb]
\begin{center}
\includegraphics[angle=0, width=14.0cm]{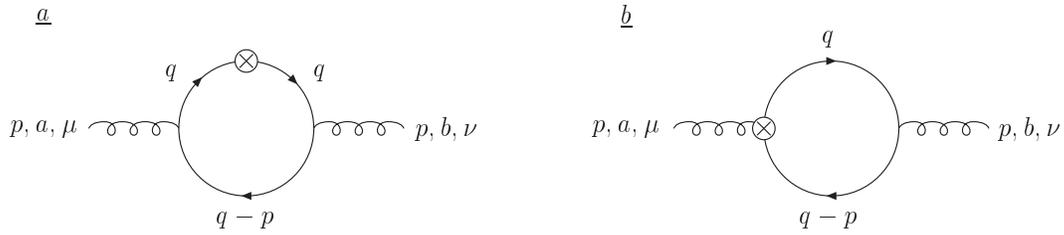} 
\end{center}
\caption{\label{aqg1diag}
\sf 
The Feynman diagrams contribution to
the operator matrix element $A_{Qg}$ at 
$O(a_s)$. Weavy lines denote gluons, and the 
full arrow lines are the heavy quark lines. The Feynman rules for the
operator insertions are given in Figure~\ref{fig:1}.
}
\end{figure}

At one--loop order, only gluonic terms contribute to heavy flavor 
production. The complete calculation of the differential scattering cross
section $d^2 \sigma(\gamma^* + N \rightarrow Q\overline{Q})/dx d Q^2$ was 
performed in Refs.~\cite{LO}. In the limit $Q^2 \gg m^2$, the heavy flavor 
Wilson coefficients are in the $\overline{\rm MS}$--scheme
  \begin{eqnarray}
  \label{eqHL}
    H_{L,g}^{(1)}\Biggl(\frac{Q^2}{m^2},\frac{m^2}{\mu^2}\Biggr)
               &=&\widehat{C}_{L,g}^{(1)}\Biggl(\frac{Q^2}{\mu^2}\Biggr)~,
                 \label{H1Lgexp} \\
  \label{eqH2}
    H_{2,g}^{(1)}\Biggl(\frac{Q^2}{m^2},\frac{m^2}{\mu^2}\Biggr)
               &=&A_{Qg}^{(1)}\Biggl(\frac{m^2}{\mu^2}\Biggr)
                  +\widehat{C}_{2,g}^{(1)}\Biggl(\frac{Q^2}{\mu^2}
                  \Biggr)~.                    \label{H12gexp}
  \end{eqnarray}
Here
$\widehat{C}_{(2,L),g}^{(1)}({Q^2}/{\mu^2})$ denote the massless one-loop
Wilson coefficients and $A_{Qg}^{(1)}({m^2}/{\mu^2})$ is the massive 
one--loop operator matrix element. Since $F_L(x,Q^2)$ is collinear finite
at leading order, the Wilson coefficient $\widehat{C}_{L,g}^{(1)}({Q^2}/{\mu^2})$ 
is universal and independent of the scheme or method of calculation.
$H_{L,g}^{(1)}({Q^2}/{m^2},{m^2}/{\mu^2})$ does therefore not 
contain contributions due to operator matrix elements at this order. The 
logarithmic contributions to $F_L(x,Q^2)$ emerge as $(m^2/Q^2) \cdot 
\ln(Q^2/m^2)$ and vanish in the limit $Q^2 \gg m^2$. The massive operator 
matrix element $A_{Qg}^{(1)}({m^2}/{Q^2})$ is calculated from the
diagrams in Figure~\ref{aqg1diag}. 
The symbol $\otimes$ denotes the 
operator insertion, cf.~Figure~\ref{fig:1}. The massive operator matrix 
elements are obtained by contracting the diagrams with the projector
\begin{eqnarray}
\hat{A}_{Qg}\left(\varepsilon,\frac{m^2}{\mu^2}, a_s\right) = 
-\frac{1}{N_c^2-1}\frac{1}{D-2} g_{\mu\nu} 
\delta^{ab} (\Delta . p)^{-N}  G_{Q,\mu\nu}^{ab} 
\end{eqnarray}
where $a$ and $b$ are the outer color indices and $\mu$ and $\nu$ are the 
Lorentz-indices and $G_{ij}$ is the respective Green's function.
The diagrams yield 
   \begin{eqnarray}        
    A^{Qg}_{a}&=&
       -\frac{1+(-1)^N}{2}
8a_sT_RS_{\ep}\Biggl(\frac{m^2}{\mu^2}\Biggr)^{\ep/2} 
       \frac{1}{(2+\ep)\ep}
        \exp\Biggl\{\sum_{l=2}^{\infty}\frac{\zeta_l}{l}
        \Biggl(\frac{\varepsilon}{2}\Biggr)^l\Biggr\} 
        \N\\
     && \times \frac{2(N^2+3N+2)+\ep(N^2+N+2)}{N(N+1)(N+2)}~,\\
    A^{Qg}_{b}&=&
       32 \frac{1+(-1)^N}{2}
     a_sT_RS_{\ep}\Biggl(\frac{m^2}{\mu^2}\Biggr)^{\ep/2} 
       \frac{1}{(2+\ep)\ep}
        \exp\Biggl\{\sum_{l=2}^{\infty}\frac{\zeta_l}{l}
        \Biggl(\frac{\varepsilon}{2}\Biggr)^l\Biggr\}
        \frac{1}{(N+1)(N+2)}~.
   \end{eqnarray}
For the unpolarized operator matrix elements only the even moments 
contribute due to the current crossing relations, see e.g. 
\cite{JBNK}. The analytic continuation is performed starting from the even 
moments. 
The matrix elements  have to be expanded to $O(\varepsilon)$, since these 
terms are 
needed for the 2--loop result:
    \begin{eqnarray}
     \hat{A}^{(1)}_{Qg}&=&\frac{1}{a_s}\Biggl[
                          A^{Qg}_{a}+A^{Qg}_{b}\Biggr] \N\\
                       &=&-S_{\ep}T_R
                          \Biggl(\frac{m^2}{\mu^2}\Biggr)^{\ep} 
                          \frac{1}{\ep}
                          \exp\Biggl\{\sum_{l=2}^{\infty}\frac{\zeta_l}
                          {l}\Biggl(\frac{\varepsilon}{2}\Biggr)^l\Biggr\}
                          \frac{8(N^2+N+2)}{N(N+1)(N+2)} \N\\
                       &=&S_{\ep}T_R
                          \Biggl(\frac{m^2}{\mu^2}\Biggr)^{\ep} 
                          \Biggl(-\frac{1}{\ep}-\frac{\zeta_2}{8}\ep\Biggr)
                          \frac{8(N^2+N+2)}{N(N+1)(N+2)} + O(\varepsilon^2)~.\label{OME1LUN}
    \end{eqnarray}
Note that the term $O(\varepsilon^0)$ vanishes. In $x$--space the result
reads
    \begin{eqnarray}
     \hat{A}_{Qg}^{(1)}&=&S_{\varepsilon}\Biggl(\frac{m^2}
                 {\mu^2}\Biggr)^{\varepsilon/2}\Biggl[-\frac{1}{\varepsilon}
            \widehat{P}_{qg}^{(0)}(x)
            +a_{Qg}^{(1)}+\varepsilon\overline{a}_{Qg}^{(1)}\Biggr]~,
    \end{eqnarray}
     where 
    \begin{eqnarray}
     a_{Qg}^{(1)}           &=&0~,\\
     \overline{a}_{Qg}^{(1)}&=&-\frac{\zeta_2}{8}\widehat{P}^{(0)}_{qg}(x)~,
    \end{eqnarray}
and $\widehat{P}_{qg}^{(0)}(x)$ 
denotes the leading order splitting function for the transition $g 
\rightarrow q$ for one (heavy) flavor. 
At one--loop order, the renormalized
operator matrix element $A_{Qg}^{(1)}$ is obtained by  
\begin{eqnarray}
A_{Qg}^{(1)}\left(\frac{m^2}{\mu^2}\right) = \hat{A}_{Qg}^{(1)} + Z_{O,Qg}^{-1,(1)}~,
\end{eqnarray}
where
\begin{eqnarray}
Z_{O,Qg}^{(1)} =  S_\varepsilon \left[-\frac{1}{\varepsilon} P_{qg}^{(0)}\right]~.
\end{eqnarray}
Here, $Z_{Qg}^{(1)}$ removes the ultraviolet singularities. At this order  no collinear singularities are 
present. Choosing $\mu^2 = Q^2$ the massless Wilson 1--loop coefficients in the $\overline{\rm MS}$--scheme 
\cite{LWILS1} are given by~:
\begin{eqnarray}
\hat{C}_{2,g}^{(1)}(z) &=&  \left\{\frac{1}{2} \widehat{P}_{qg}^{(0)}(z) \left[ \ln 
\left(\frac{1-z}{z}\right) -4 \right] + 12 T_R\right\}~,\\
\hat{C}_{L,g}^{(1)}(z) &=& 16 T_R z(1-z)~. 
\end{eqnarray}
The asymptotic heavy quark Wilson coefficients (\ref{eqHL},\ref{eqH2}) are
   \begin{eqnarray}
    H_{L,g}^{(1)}\Biggl(z,\frac{Q^2}{m^2}\Biggr)
    &=& 16 T_R z(1-z)~,
    \\
    H_{2,g}^{(1)}\Biggl(z,\frac{Q^2}{m^2}\Biggr)
    &=& T_R\left\{
        4(z^2+(1-z)^2) \left[ \ln\left(\frac{1-z}{z}\right) + \ln\left(\frac{Q^2}{m^2}\right)\right]
    - 32 z^2 + 32 z - 4\right\}~.
\nonumber\\
   \end{eqnarray}
These are precisely the expressions  obtained in the limit $m^2/Q^2 
\rightarrow 0$ from
\cite{LO} up to the constant term $(m^2/Q^2)^0$~:
   \begin{eqnarray}
    H_{L,g}^{{\rm compl},(1)}\Biggl(z,\frac{Q^2}{m^2}\Biggr)
    &=& 16 T_R \left[v z(1-z) + 2 \frac{m^2}{Q^2} z^2 \ln\left(\frac{1-v}{1+v}\right)\right]~,
    \\
    H_{2,g}^{{\rm compl},(1)}\Biggl(z,\frac{Q^2}{m^2}\Biggr)
    &=& 8 T_R\left\{ v \left[-\frac{1}{2} + 4z - 4 z^2 - 2 \frac{m^2}{Q^2} z(1-z)\right]
\right. \nonumber\\ & &\left.+ \left[ -\frac{1}{2} +z -z^2 + 2 \frac{m^2}{Q^2} (3 z^2 -z) + 4 \frac{m^4}{Q^4} 
z^2 \right]
\ln\left(\frac{1-v}{1+v}\right) \right\}~,
   \end{eqnarray}
where $v$ denotes the cms-velocity of the heavy quarks,
\begin{eqnarray}
v = \sqrt{1- \frac{4 m^2}{Q^2} \frac{z}{1-z}}~.
\end{eqnarray}
\section{The two-loop massive operator matrix elements}

\begin{figure}[h]
\begin{center}
\includegraphics[angle=0, width=13.0cm]{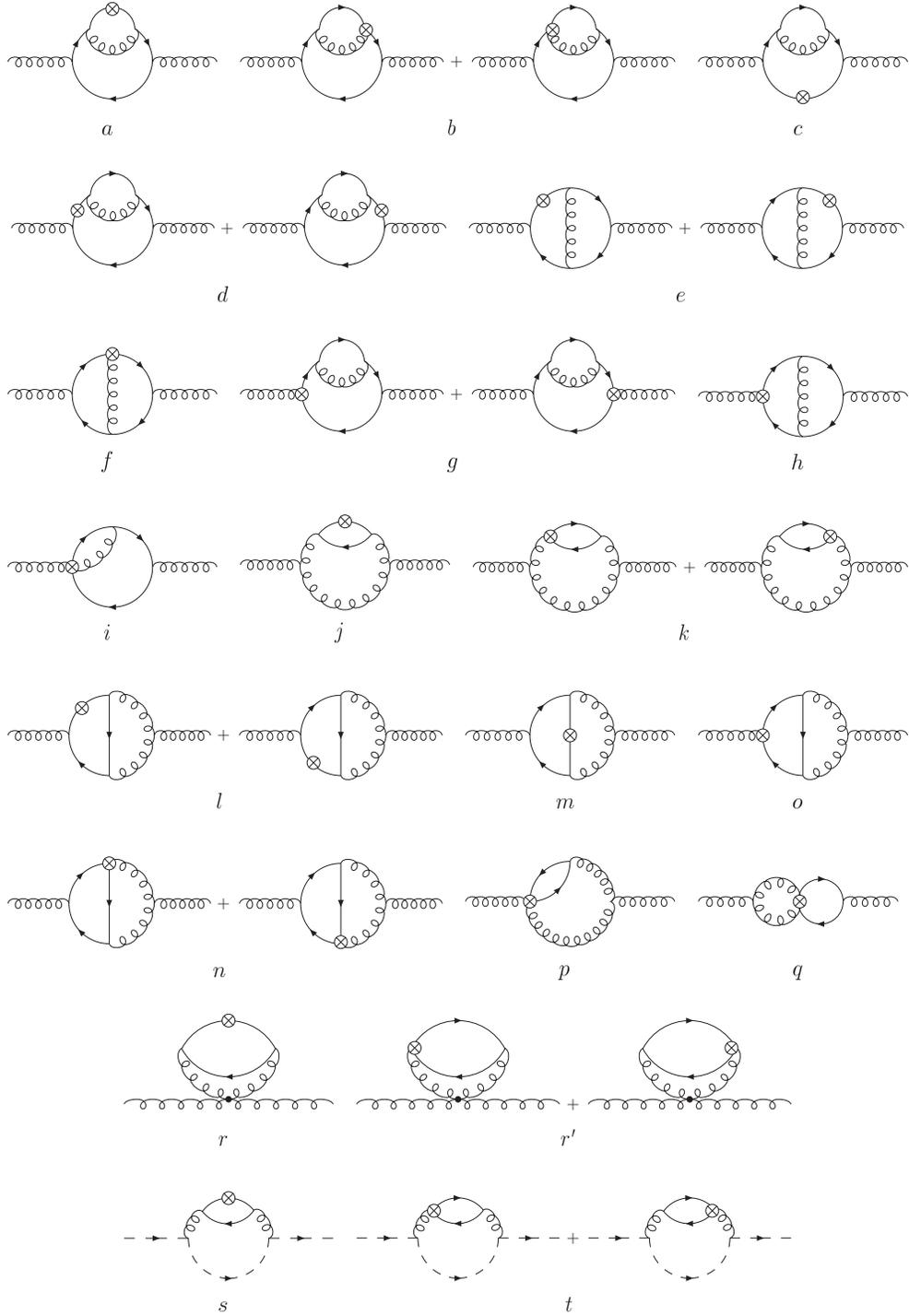} 
\end{center}
\caption{\label{fig:3}
\sf The diagrams contributing to the operator matrix element $A_{Qg}$ at 
$O(a_s^2)$. Weavy lines denote gluons, dashed lines ghosts, and the 
full arrow lines are the heavy quark lines. The Feynman rules for the
operator insertions are given in Figure~\ref{fig:1}.
}
\end{figure}

\begin{figure}[h]
\begin{center}
\includegraphics[angle=0, width=13.0cm]{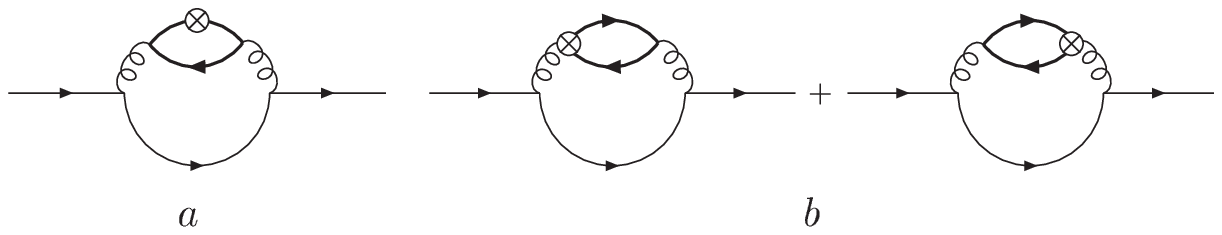} 
\end{center}
\caption{\label{fig:4}
\sf The diagrams contributing to the operator matrix element 
$A_{Qq}^{\rm PS}$ at 
$O(a_s^2)$. Weavy lines denote gluons, the 
thick full arrow lines are the heavy quark lines, and the thin full lines 
are light quark lines. The Feynman rules for 
the
operator insertions are given in Figure~\ref{fig:1}.
}
\end{figure}

\begin{figure}[h]
\begin{center}
\includegraphics[angle=0, width=15.0cm]{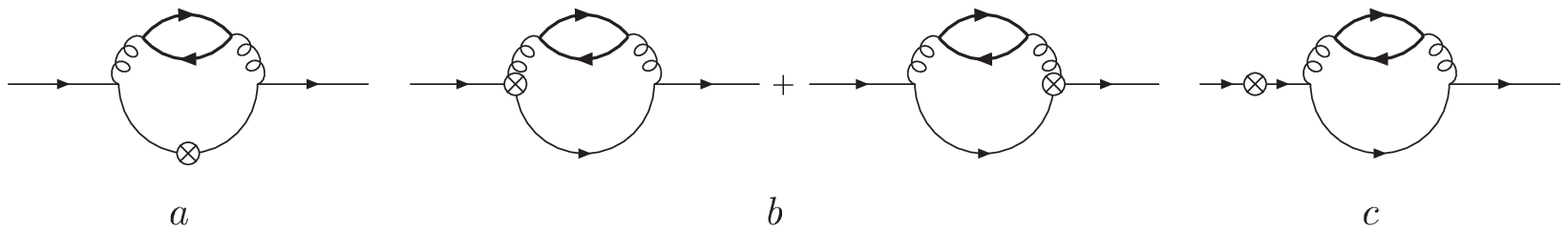} 
\end{center}
\caption{\label{fig:5}
\sf The diagrams contributing to the operator matrix element 
$A_{Qqq}^{\rm NS}$ at 
$O(a_s^2)$. Weavy lines denote gluons, the 
full arrow lines are the heavy quark lines, and the thin full lines are 
light quark lines. The Feynman rules for the
operator insertions are given in Figure~\ref{fig:1}.
}
\end{figure}

\vspace{1mm}\noindent
There are three classes of 2--loop contributions to the massive operator 
matrix elements: the gluonic contributions (diagrams Figure~\ref{fig:3}), the pure-singlet 
contributions (diagrams Figure~\ref{fig:4}) and the non--singlet contributions  
(diagrams Figure~\ref{fig:5}). The diagrams are either one-loop insertions into one-loop diagrams
or, in case of the gluonic contributions, also  genuine two--loop diagrams.
The calculation is performed using {\tt FORM} \cite{FORM} and {\tt Maple} procedures. 
We calculate all diagrams directly, i.e. without decomposing them using the
integration-by-parts method as done in \cite{Buza:1995ie}. 
The integrals to be performed are more involved. However,
we avoid a large proliferation of terms in the results, in our case of nested  sums of different kind, which 
add up to zero. Even in the case of individual diagrams, which are calculated in Feynman gauge, 
only a very small number of harmonic sums contributes finally. One of these, $S_{2,1}(N)$, does
not emerge in the operator matrix elements.
The results for the individual non-renormalized diagrams are given in Appendix~A. 
It turns out, that some of the diagrams can be easily calculated to all orders
in $\varepsilon$. 
The diagrams can be represented in terms of linear combinations of generalized 
hypergeometric functions~\cite{HGF}. Due to the given topologies the most complex function is 
$_3F_2(a_1,a_2,a_3;b_1,b_2;1)$. In these representations, the conformal mapping of 
Feynman parameters is essential. The scalar integrals associated to the genuine 
two--loop diagrams have also been calculated  using the Mellin--Barnes technique \cite{MB1,MB2}, 
cf.~Refs.~\cite{Bierenbaum:2006mq,Bierenbaum:2007dm}, and checked for fixed moments using the package {\tt 
MB} \cite{Czakon:2005rk}. The $\varepsilon$--expansion can be performed prior to the summation. It  
results into finite and infinite sums of various types, including harmonic sums attached with
Euler Beta-functions and binomials. Here, we face a more general situation than in massless
calculations, despite the fact that we work in the limit $Q^2 \gg m^2$. The sums are given in 
Appendix~B. They were performed using suitable integral representations or using difference equations.
Finally, the results
depend only on harmonic sums, which are reduced further applying their algebraic relations
\cite{ALGEBRA}.

After mass renormalization, the operator matrix element $\hat{A}^{(2)}_{Qg}\Bigl({m^2}/{\mu^2},\ep\Bigr)$ is 
given by
     \begin{eqnarray}
      \hat A_{Qg}^{(2)} &=&S_\ep^2 \Big( \frac{m^2}{\mu^2}\Big)^\ep
                              \Big[ \frac{1}{\ep^2} \Big\{\frac{1}{2} 
                              \hat{P}^{(0)}_{qg} \otimes ( P^{(0)}_{qq} - 
                              P^{(0)}_{gg})+ \beta_0 \hat{P}^{(0)}_{qg} \Big\}
                             + \frac{1}{\ep} \Big\{ - \frac{1}{2} 
                              \hat{P}^{(1)}_{qg}
                               \Big\} + a^{(2)}_{Qg} \Big] \N\\
                        &&       - \frac{2}{\ep} S_\ep \beta_{0,Q}
                            \sum_{N_H=4}^6\Big( \frac{m_{N_H}^2}{\mu^2} \Big)^{\ep/2}
                             \Big( 1 + \frac{\ep^2}{8} \zeta_2 \Big)
                              \hat A^{(1)}_{Qg}~. \label{AQg2f}
\nonumber\\
&=&
     S_{\ep}^2 \left(\frac{m^2}{\mu^2}\right)^{\ep}\Biggl(
         \frac{1}{\ep^2}\Biggl\{ 
           8T_RC_F\frac{N^2+N+2}{N(N+1)(N+2)}
                  \Biggl[
                        -4S_1(N)
                         +\frac{3N^2+3N+2}{N(N+1)}  
                   \Biggr] \N\\
        &&+32T_RC_A\frac{N^2+N+2}{N(N+1)(N+2)}
                  \Biggl[
                         S_1(N)
                         -2\frac{N^2+N+1}{(N-1)N(N+1)(N+2)}
                  \Biggr]
       \Biggr\} \N\\
    &&+ \frac{1}{\ep}
        \Biggl\{
          4T_RC_F
               \Biggl[
                2\frac{N^2+N+2}{N(N+1)(N+2)}
                     \Bigl(
                      S_2(N)-S^2_1(N)
                     \Bigr)
               +4\frac{S_1(N)}{N^2}  \N\\
    &&         -\frac{5N^6+15N^5+36N^4+51N^3+25N^2+8N+4}
                      {N^3(N+1)^3(N+2)}
               \Biggr] \N\\
    &&   +T_RC_A 
               \Biggl[ 
                8\frac{N^2+N+2}{N(N+1)(N+2)}
                     \Bigl(             
                     -2\beta'(N+1)
                     +S_2(N)
                     +S^2_1(N)
                     -\zeta_2
                     \Bigr)              \N
\end{eqnarray} \begin{eqnarray}
    &&          -32\frac{2N+3}{(N+1)^2(N+2)^2}S_1(N)
                -8\frac{\hat{P}_1(N)}
                       {(N-1)N^3(N+1)^3(2+N)^3}
               \Biggr]
        \Biggr\}  \N\\
    && +a_{Qg}^{(2)}(N)\Biggr) 
       +S_{\ep} 
        \frac{8}{3\ep}T_R
        \Bigl(1+\frac{\zeta_2}{8}\ep^2
        \Bigr)\sum_{N_H=4}^6\Biggl(\frac{m_{N_H}^2}{\mu^2}\Biggr)^{\ep/2}
        \hat{A}^{(1)}_{Qg}~. \label{AQg2fres}
     \end{eqnarray}  
For the ghost contributions in Figure~\ref{fig:3}, the projector reads
\begin{eqnarray}
\hat{A}_{Qg}^{(2),{\rm ghost}}\left(\varepsilon, \frac{m^2}{\mu^2}, 
a_s\right) = \frac{1}{N_c^2-1} \frac{1}{D-2} \delta^{ab} 
\left(\Delta.p\right)^{-N} G_Q^{ab,(2),{\rm ghost}}~. 
\end{eqnarray}
From the terms $\propto 1/\varepsilon^2, 1/\varepsilon$, one may 
determine the respective QCD splitting functions to two--loop order (\ref{eqPqq0}--\ref{eqPqg1}), 
which are recalculated in this way. 

The constant term in $\hat{A}^{(2)}_{Qg}\Bigl({m^2}/{\mu^2},\ep\Bigr)$ reads~:
     \begin{eqnarray}
       a_{Qg}^{(2)}(N) &=& 4 C_F T_R \Biggl\{
                 \frac{N^2+N+2}{N\left(N+1\right)\left(N+2\right)}
                                         \Biggl[
                         -\frac{1}{3}S_1^3(N-1)+\frac{4}{3}S_3(N-1) 
                           \N\\ & & \hspace{1.5cm}
                         -S_1(N-1)S_2(N-1)-2\zeta_2S_1(N-1)
                                          \Biggr]
                  + \frac{2}{N(N+1)}S_1^2(N-1)
                   \nonumber\\
& & \hspace{1.5cm}
                  +  \frac{N^{4}+16\,{N}^{3}+15\,{N}^{2}-8\,N-4}
                     {N^2\left(N+1\right)^{2}\left(N+2\right)}S_2(N-1)
\nonumber\\
                  & & \hspace{1.5cm}
                  +\frac {3\,{N}^{4}+2\,{N}^{3}+3\,{N}^{2}-4\,N-4}
                    {2 N^2 \left( N+1 \right) ^{2} \left( N+2 \right) } \zeta_2 
\nonumber\\ 
                  & & \hspace{1.5cm}
                  +\frac {N^4-N^3-16 N^2 + 2N +4}
                    {N^2 \left( N+1 \right) ^{2}\left( N+2 \right)} S_1(N-1)
                  + \frac {\hat{P}_2(N)}
                  {2 N^4 \left( N+1 \right) ^{4} \left( N+2 \right) }
                                         \Biggr\}  \N
\\
        & & +4 C_A T_R\Biggl\{ \frac{N^2+N+2}{N(N+1)(N+2)} 
          \Biggl[ 4 \Mvec\left[\frac{\Li_2(x)}{1+x}\right](N+1) +\frac{1}{3} 
           S_1^3(N)  +3 S_2(N) S_1(N) \nonumber\\ 
         & & \hspace{1.5cm} 
            + \frac{8}{3} S_3(N) 
         + \beta''(N+1) - 4 \beta'(N+1) S_1(N) - 4 \beta(N+1) 
          \zeta_2 
          +\zeta_3\Biggr]\nonumber\\
           & & \hspace{1.5cm}
    -\,{\frac {{N}^{3}+8\,{N}^{2}+11\,N+2}{N \left( N+1 \right) ^{2} \left( N+2
            \right) ^{2}}} S_1^2(N)
           -2\,{\frac {N^4 - 2 N^3 + 5 N^2+ 2 N + 2} 
          { \left( N-1 \right)  N^2 \left( N+1 \right) ^{2} \left( N+2 
            \right) }} 
            \zeta_2\nonumber 
         \end{eqnarray}
         \begin{eqnarray}
            & & \hspace{1.5cm}
         -\, \frac {7 N^5 + 21 N^4 + 13 N^3 + 21 N^2 +18 N +16} 
              { (N-1) N^2 \left( N+1 \right) ^{2} \left( N+2 \right) ^{2}} 
            S_2(N)  \N\\
         & & \hspace{1.5cm}
          -\,{\frac {{N}^{6}
          +8\,{N}^{5}
          +23\,{N}^{4}
          +54\,{N}^{3}
          +94\,{N}^{2}
          +72\,N
          +8}{ N \left( N+1 \right) ^
          {3} \left( N+2 \right) ^{3} }} S_1(N)
          \nonumber\\ 
           & & \hspace{1.5cm}
           -4 \, \frac { \left( {N}^{2} - N -4 \right)}
              { \left( N+1 \right) ^{2} \left( N+2 \right) ^{2}} \beta'(N+1)
           + \frac{\hat{P}_3(N)}{(N-1) N^4 (N+1)^4 (N+2)^4}\Biggr\}~.
         \label{eqaQg2}
     \end{eqnarray}
     The polynomials in Eqs. (\ref{AQg2fres},~\ref{eqaQg2}) read
     \begin{eqnarray}
          \hat{P}_1(N)&=&\!N^9\!+6N^8\!+15N^7\!+25N^6\!+36N^5\!+85N^4\!+128N^3+104N^2+64N+16~, \\
          \hat{P}_2(N) &=& 12 N^{8}
                  +54 N^{7} 
                 +136 N^{6}
                 +218 N^{5}
                 +221 N^{4}
                 +110 N^{3}
                   -3 N^{2}
                  -24 N
                   -4~,   \\
          \hat{P}_3(N) &=& 2\,{N}^{12}
                +20\,{N}^{11}
                +86\,{N}^{10}
               +192\,{N}^{9}
               +199\,{N}^{8}
                    -{N}^{7}
               -297\,{N}^{6}
               -495\,{N}^{5}\nonumber\\ & &
               -514\,{N}^{4} 
               -488\,{N}^{3}
               -416\,{N}^{2}
               -176\,N
                -32~.
     \end{eqnarray}

The pure--singlet 
$\hat{A}_{Qq}^{{\rm PS},(2)}\Bigl({m^2}/{\mu^2},\ep\Bigl)$
operator matrix element is given by
     \begin{eqnarray}
      \hat{A}_{Qq}^{{\rm PS},(2)}\Bigl(\frac{m^2}{\mu^2},\ep\Bigl)
                        &=& 
                          S_\ep^2 \Big( \frac{m^2}{\mu^2} \Big)^{\ep}
                         \Big[ \frac{1}{\ep^2} \Big\{-\frac{1}{2} 
                           \hat{P}^{(0)}_{qg}\otimes P^{(0)}_{gq} \Big\}
                         + \frac{1}{\ep} \Big\{
                        -\frac{1}{2} \hat{P}^{{\rm PS},(1)}_{qq}  \Big\}
                      + a^{{\rm PS}, (2)}_{Qq} \Big]~. \label{AQq2f}
\nonumber\\ 
\!\!          &=&S_{\ep}^2\Biggl(\frac{m^2}{\mu^2}\Biggr)^{\ep}
                \Biggl\{
             -\frac{1}{\ep^2}T_RC_F
              \frac{16(N^2+N+2)^2}{(N-1)N^2(N+1)^2(N+2)} \N\\
\!\!           &-&\frac{1}{\ep}T_RC_F\frac{8(5N^5+32N^4+49N^3+38N^2+28N+8)}
                                {(N-1)N^3(N+1)^3(N+2)^2}
              +a_{Qq}^{{\rm PS},(2)}(N) 
              \Biggr\}~. \label{APS2Qq}
     \end{eqnarray}
In this case and the non--singlet case, the projector
\begin{eqnarray}
\hat{A}_{Qq;(qq,Q)}^{(2)}\left(\varepsilon, \frac{m^2}{\mu^2}, 
a_s\right) = \frac{1}{N_c} \delta^{ij} \frac{1}{4} 
\left(\Delta.p\right)^{-N} {\rm Tr}\left[p\negthickspace\! / 
G_{Q,(q)}^{ij,(2)}\right]~. 
\end{eqnarray}
is applied. Here $i$ and $j$ denote the matrix-elements of the Gell-Mann 
matrices.
The constant term is obtained by
     \begin{eqnarray}
      a_{Qq}^{{\rm PS},(2)}(N)&=&T_RC_F\Biggl\{
      -4\frac{(N^2+N+2)^2}
             {(N-1)N^2(N+1)^2(N+2)}\left(2S_2(N)+\zeta_2\right)\N\\
      &&+\frac{4\hat{P}_4(N)}{(N-1)N^4(N+1)^4(N+2)^3} \Biggr\},  \label{aPS2Qq}\\
      \hat{P}_4(N)&=&N^{10}+8N^9+29N^8+49N^7-11N^6-131N^5-161N^4\\
            &&-160N^3-168N^2-80N  -16~.\N
      \end{eqnarray} 

Finally, the non-singlet operator matrix element 
     $\hat{A}^{{\rm NS},(2)}_{qq,Q}\Bigl({m^2}/{\mu^2},\ep\Bigr)$ reads
      \begin{eqnarray}
      \hat{A}^{{\rm NS},(2)}_{qq,Q}\Bigl(\frac{m^2}{\mu^2},\ep\Bigr)
                         &=&S_\ep^2 \Big( \frac{m^2}{\mu^2} 
                            \Big)^\ep \Big[ \frac{1}{\ep^2}
                            \Bigl\{ - \beta_{0,Q} P^{(0)}_{qq} \Bigr\}
                            + \frac{1}{\ep}
                          \Bigl\{ -\frac{1}{2} P^{{\rm NS},(1)}_{qq,Q}\Bigr\}
                           +a^{{\rm NS},(2)}_{qq,Q}\Big] \,. \label{AqqQ2f}
     \nonumber\\
        &=&S_{\varepsilon}^2\Biggl(\frac{m^2}{\mu^2}\Biggr)^{\varepsilon}
            \Biggl\{
              \frac{1}{\ep^2}T_RC_F\Biggl[
                                  -\frac{32}{3}S_1(N)
                                  +8\frac{3N^2+3N+2}{3N(N+1)}
                                   \Biggr] \N\\
         &&   +\frac{1}{\ep}T_RC_F\Biggl[
                             \frac{16}{3}S_2(N) 
                            -\frac{80}{9}S_1(N) 
                            +2\frac{3N^4+6N^3+47N^2+20N-12}
                                   {9N^2(N+1)^2}
                                \Biggr]\N\\ && + a_{qq,Q}^{{\rm NS},(2)}(N) 
                         \Biggr\}\label{ANS2qqQ}~.
     \end{eqnarray}
The constant term is given by
     \begin{eqnarray}
      a_{qq,Q}^{{\rm NS},(2)}(N)&=&C_FT_R\Biggl\{
                        -\frac{8}{3}S_3(N)
                        -\frac{8}{3}\zeta_2S_1(N)
                        +\frac{40}{9}S_2(N)
                        +2\frac{3N^2+3N+2}
                               {3N(N+1)}\zeta_2
                        -\frac{224}{27}S_1(N) \N\\
                     && +\frac{219N^6+657N^5+1193N^4+763N^3-40N^2-48N+72}
                              {54N^3(N+1)^3}\Biggr\} \label{aNS2qqQ}.
     \end{eqnarray}
These results obtained in Mellin--space agree with those given in Ref.~\cite{Buza:1995ie} in 
$x$--space, cf.~\cite{Blumlein:2006mh}.   

The method applied here allowed to compactify the representation for the heavy flavor
matrix elements and Wilson coefficients. As shown in Appendix~A the individual Feynman diagrams
depend on the harmonic sums 
$S_1(N), S_2(N), S_3(N), S_{-2}(N), S_{-3}(N), S_{2,1}(N), S_{-2,1}(N)$ only. In the final result
the sum $S_{2,1}(N)$ drops out.
The $x$-space representation in \cite{Buza:1995ie} contains the following 48 functions~:
\[
\renewcommand{\arraystretch}{1.0}
\begin{array}{ccccc}
\delta(1-x)          & 1                   & \ln(x)              & \ln^2(x)            & \ln^3(x)            
\\ 
& & & & \\
\ln(1-x)             & \ln^2(1-x)          & \ln^3(1-x)          & \ln(x) \ln(1-x)     & \ln(x) \ln^2(1-x)   
\\ 
& & & & \\
\ln^2(x) \ln(1-x)    & \ln(1+x)            & \ln(x) \ln(1+x)     & \ln^2(x) \ln(1+x)   &  \Li_2(1-x)         
\\ 
& & & & \\
\ln(x) \Li_2(1-x)    & \ln(1-x) \Li_2(1-x) & \Li_3(1-x)        & S_{1,2}(1-x)        & S_{1,2}(-x)         
\\
& & & & \\
\ds \frac{\ds 1}{\ds 1 -x}        &
\ds \frac{\ds 1}{\ds 1 +x}        &
\ds \frac{\ds \ln(x)  }{\ds 1 -x} & 
\ds \frac{\ds \ln^2(x)}{\ds 1 -x} & 
\ds \frac{\ds \ln^3(x)}{\ds 1 -x} \\ 
& & & & \\
\ds \frac{\ds \ln(x)  }{\ds 1 +x}         & 
\ds \frac{\ds \ln^2(x)}{\ds 1 +x}         &
\ds \frac{\ds \ln^3(x)}{\ds 1 +x}         &
\ds \frac{\ds        \ln(1+x)}{\ds 1+x}   &
\ds \frac{\ds \ln(x) \ln(1+x)}{\ds 1+x}   \\
& & & & \\
\ds \frac{\ds \ln(x) \ln^2(1+x)}{\ds 1+x}   &
\ds \frac{\ds \ln^2(x) \ln(1+x)}{\ds 1+x}   &
\ds \frac{\ln(x) \ln(1-x)}{1-x}           &
\ds \frac{\ln(x) \ln^2(1-x)}{1-x}           &
\ds \frac{\ln(1-x) \Li_2(x)}{1-x}           \\
& & & & \\
\ds \frac{\ds \Li_2(1-x)}{\ds 1 -x}         &
\ds \frac{\ds \ln(x) \Li_2(1-x)}{\ds 1 -x}  &
\ds \frac{\ds \ln(x) \Li_2(1-x)}{\ds 1 +x}  &
\ds \frac{\ds \ln(1+x) \Li_2(-x)}{\ds 1 +x} & 
\ds \ln(1+x) \Li_2(-x)                      \\
& & & & \\
    \Li_2(-x)                               &
\ds \frac{\Li_2(-x)}{1+x}                   &
\ds \frac{\ln(x) \Li_2(-x)}{1+x}            &
\ds \frac{\ds \Li_3(1-x)}{\ds 1 -x}         &
\ds \frac{\ds \Li_3(-x )}{\ds 1 +x}         \\
& & & & \\
\ds \frac{\ds S_{1,2}(1-x)}{\ds 1 -x}       &
\ds \frac{\ds S_{1,2}(1-x)}{\ds 1 +x}       &
\ds \frac{\ds S_{1,2}(-x)}{\ds 1 +x}        & & \\
\end{array}
\]
As shown in \cite{HSUM1}, various of these functions have Mellin transforms which contain triple 
sums, which do not occur in our approach even on the level of individual diagrams.

\vspace{3mm}\noindent
In the Mellin--space representation, the sums listed in Table~1 contribute 
to the result of the  individual 
diagrams.
     Note, that we express single harmonic sums with negative index
     in terms of $\beta$-functions and their derivatives, cf.~\cite{HSUM1}.
     They can be traced back to the single non-alternating harmonic sums, allowing for 
     half-integer arguments. Therefore, all single harmonic sums form an equivalence class
     represented by $S_1(N)$, from which through differentiation and half-integer relations
     the other single harmonic sums are easily derived.
     Further the equality, 
     \begin{eqnarray}
      \Mvec\left[\frac{\Li_2(x)}{1+x}\right](N+1)  - \zeta_2 \beta(N+1)
      = (-1)^{N+1} \left[S_{-2,1}(N)  + \frac{5}{8} \zeta_3\right]
      ~\label{hhhh}
     \end{eqnarray}
     holds.~\footnote{We correct a typo in~\cite{Blumlein:2006mh}.
     The argument of the Mellin-transform in (\ref{eqaQg2})
     reads $N+1$, not $N$.}
     Therefore, the  operator matrix 
     element $\hat{A}^{(2)}_{Qg}$ depends on one non-trivial basic 
     function only~\cite{HSUM1}. 
     The absence of harmonic sums containing $\{-1\}$ as index was noted
     before for all other classes of (space- and time-like) anomalous dimensions
     and Wilson coefficients, including those for other hard processes having been 
     calculated so far, cf.~\cite{MATH,JB07}. This can be seen if one represents the 
     respective expressions in form of weighted harmonic sums, following an earlier suggestion of one of the 
     authors. Linear representations do not allow this since
     they are non-minimal and contain algebraic redundancies.

      \begin{center}
      \renewcommand{\arraystretch}{1.1}
      \begin{tabular}{||l|c|c|c|c|c|c|c|r||}
      \hline \hline
Diagram  & $S_1$ & $S_2$ & $S_3$ & $S_{-2}$ & $S_{-3}$ & $S_{2,1}$ & $S_{-2,1}$ & \# $x$-space 
fct.\\
      \hline \hline

A             &       & +     &       &          &          &           &            & 8              \\
B             & +     & +     & +     &          &          & +         &            & 10             \\
C             &       & +     &       &          &          &           &            & 4              \\
D             & +     & +     &       &          &          &           &            & 5              \\
E             & +     & +     &       &          &          &           &            & 9              \\
F             & +     & +     & +     &          &          & +         &            & 24             \\
G             & +     & +     &       &          &          &           &            & 6              \\
H             & +     & +     &       &          &          &           &            & 7              \\
I             & +     & +     & +     & +        & +        & +         & +          & 20             \\
J             &       & +     &       &          &          &           &            & 7              \\
K             &       & +     &       &          &          &           &            & 7              \\
L             & +     & +     & +     &          &          & +         &            & 13             \\
M             &       & +     &       &          &          &           &            & 7              \\
N             & +     & +     & +     & +        & +        & +         & +          & 38             \\
O             & +     & +     & +     &          &          & +         &            & 13             \\
P             & +     & +     & +     &          &          & +         &            & 14             \\
\hline
S             &       & +     &       &          &          &           &            & 7              \\
T             &       & +     &       &          &          &           &            & 7              \\
\hline
${\rm PS}_a$  &       & +     &       &          &          &           &            &                \\
${\rm PS}_b$  &       & +     &       &          &          &           &            & 7              \\
\hline
${\rm NS}_a$  &       &       &       &          &          &           &            &                \\
${\rm NS}_b$  & +     & +     & +     &          &          &           &            & 5              \\
      \hline \hline
$\Sigma$      & +     & +     & +     & +        & +        &           & +          & 48             \\
\hline \hline
      \end{tabular}
      \renewcommand{\arraystretch}{1.0}
      \end{center}
     \noindent

\vspace{3mm}\noindent
{\sf Table~1:~Harmonic sums contributing to the individual diagrams compared to the number of functions
in $x$--space, Ref.~\cite{Buza:1995ie}.}

\vspace{2mm}

The expressions for the renormalized two--loop operator matrix elements (\ref{eqOM}) are given by, cf. 
\cite{Buza:1995ie},
\begin{eqnarray}
\label{eqA1}
A_{Qg}^{(2)}\left(\frac{m^2}{\mu^2},N\right) &=&
\left\{
\frac{1}{8} P_{qg}^{(0)}(N) 
\left[P_{qq}^{(0)}(N) - P_{gg}^{(0)}(N) \right] 
+ \frac{1}{4} \beta_0 P_{qg}^{(0)}(N) \right\} \ln^2\left(\frac{m^2}{\mu^2}\right)
\nonumber\\ & &
-\frac{1}{2} P_{qg}^{(1)}(N) \ln\left(\frac{m^2}{\mu^2}\right) + a_{Qg}^{(2)}(N)
\nonumber\\ & & + \overline{a}_{Qg}^{(1)}(N)\left\{ 2 \beta_0 +
\left[P_{qq}^{(0)}(N) - P_{gg}^{(0)}(N) \right] \right\}
\\
A_{Qq}^{{\rm PS},(2)}\left(\frac{m^2}{\mu^2},N\right) &=&
\left\{-\frac{1}{8}  
P_{qg}^{(0)}(N) P_{gq}^{(0)}(N) \right\} \ln^2\left(\frac{m^2}{\mu^2}\right) 
+ \left\{-\frac{1}{2} P_{qq}^{{\rm PS},(1)}(N)\right\} \ln\left(\frac{m^2}{\mu^2}\right)
\nonumber\\ & & 
+ a_{Qq}^{{\rm PS},(2)}(N) - \overline{a}^{(1)}_{Qg}(N) P_{gq}^{(0)}(N)
\end{eqnarray}\begin{eqnarray}
\label{eqA2}
A_{qq,Q}^{{\rm NS},(2)}\left(\frac{m^2}{\mu^2},N\right) &=&
\left\{-\frac{1}{4} \beta_{0,Q} P_{qq}^{(0)}(N) \right\} \ln^2\left(\frac{m^2}{\mu^2}\right) 
+ \left\{-\frac{1}{2} P_{qq,Q}^{{\rm NS},(1)}(N)\right\} \ln\left(\frac{m^2}{\mu^2}\right)
\nonumber\\ & &
+ a_{qq,Q}^{{\rm NS},(2)}(N) + \frac{1}{4} \beta_{0,Q} \zeta_2 P_{qq}^{(0)}(N)~.
\end{eqnarray}
Besides the splitting functions up to next-to-leading order, the constant
terms $a_{ij}^{A}(N)$ and $\overline{a}_{Qg}^{(1)}(N)$ determine the
massive operator matrix elements (\ref{eqA1}--\ref{eqA2}).

The asymptotic heavy flavor Wilson coefficients $H_{2,L}(N,Q^2)$ are then given
by (\ref{eqH2g}--\ref{eqHLqNS}). The corresponding expressions in $x$--space
are obtained applying the inverse Mellin transform. Analytic continuations of the
corresponding basic functions to complex values of $N$ are given at high precision in
\cite{ANCONT1,JB07}. The inverse Mellin transform to obtain the respective 
contributions for the structure functions is  performed by a 
single precise numeric contour integral around the singularities of the problem, after
convoluting with the evolved parton densities~\cite{EVOL}. 
\section{Conclusions}
     \label{sec-conc}
     \renewcommand{\theequation}{\thesection.\arabic{equation}}
     \setcounter{equation}{0}

\vspace{1mm}\noindent
We calculated the unpolarized massive 2--loop operator matrix elements, 
which are used to express the heavy flavor Wilson coefficients in the 
asymptotic region $Q^2 \gg m^2$ for $F_2(x,Q^2)$ to $O(a_s^2)$ and for 
$F_L(x,Q^2)$ to $O(a_s^3)$. We confirm the results obtained in 
Ref.~\cite{Buza:1995ie}. The method applied in the present paper is widely 
different from the one used in \cite{Buza:1995ie}. We calculated the 
Feynman diagrams without applying the integration-by-parts method and 
worked in Mellin--space, to obey the natural symmetry of the problem.
The calculation refers to nested sums in the first place, while in \cite{Buza:1995ie}
the Feynman--parameter integrals were mapped to a single Mellin transform successively
integrating Nielsen-type integrals.
Furthermore we applied the algebraic relations between the harmonic sums to simplify the expressions
further.
The representation obtained for the individual Feynman diagrams was 
much more compact. Only a few harmonic sums contribute, which 
furthermore can be grouped into only two equivalence classes. This is to 
be compared to 48 functions in $x$--space, which were needed to express 
the result in \cite{Buza:1995ie}. The present problem exhibits a more 
involved nesting if compared to massless two-loop calculations, since the 
heavy quark mass connects Feynman parameters, although we work in the 
limit $Q^2 \gg m^2$. The representation of the Feynman-parameter integrals 
of the loop-diagrams in terms of higher transcendental functions, here 
generalized hypergeometric functions, before carrying out the 
$\varepsilon$--expansion, proved to be essential for the compactification.
In the present calculation new types of finite and infinite sums beyond 
the case of multiple harmonic sums had to be performed. The final results
could again be expressed by harmonic sums.

\newpage
\section{Appendix A: Results for the Individual 2--loop Diagrams}
     \label{sec-res}
     \renewcommand{\theequation}{\thesection.\arabic{equation}}
     \setcounter{equation}{0}
In the following, we list the results for the individual Feynman diagrams,
in some cases to all orders in $\varepsilon$, to demonstrate the simplicity
of their structure as obtained by the present method of direct calculation. 
An overall factor $\hat{a}_s^2S_{\ep}^2({\hat{m}^2}/{\mu_0^2})^{\ep}$ 
has been taken out. Here $\mu_0$ denotes the initial scale to define the strong 
coupling constant. 

\vspace{1mm}\noindent
The individual contributions to the operator matrix elements for the diagrams of Figure~\ref{fig:3} are~:
\begin{eqnarray}
A^{Qg}_a &=&
         T_R C_F\frac{-\pi}{\sin((1+\ep/2)\pi)}
         \exp\Bigl(\sum_{i=2}^{\infty}\frac{\zeta_i}{i}\ep^i\Bigr)
         \frac{\Gamma(N-\ep/2)\Gamma(N)}
         {\Gamma(N+2+\ep/2)\Gamma(N+3-\ep)}
         \frac{B(N,3)}{\ep(\ep+2)} \N\\
 &&      \Bigl(
                  16N(N+1)^2(N+2)^2 \N\\
 &&              +8(N+1)(N+2)(3N^3-N^2-6N-4)\ep \N\\
 &&              +4N(9N^4+12N^3-9N^2-28N-20)\ep^2 \N\\
 &&              +(10N^5+8N^4+6N^3+24N^2+72N+64)\ep^3 \N\\
 &&              +(2N^5-10N^4-36N^3-24N^2+24N+16)\ep^4 \N\\
 &&              +(-4N^4-4N^3+2N^2+2N-12)\ep^5 \N\\
 &&              +(2N^3+4N^2+2N-4)\ep^6
         \Bigr) \N\\
&=&
T_RC_F\Biggl\{
            \frac{1}{\ep^{2}}\frac{16}{{N}^{2}(N+1)}
           +\frac{8}{\ep}\frac{2N^3-N-2}{N^3(N+1)^2(N+2)}
           +\frac{8}{{N}^{2}(N+1)}S_2(N)
\N\\
         &&+\frac{4}{N^2(N+1)}\zeta_2
           +\frac{4P_1(N)}
              {N^4(N+1)^3(N+2)^2}
        \Biggr\} + O(\varepsilon)~,       \label{resA} \\
P_{1}(N)&=&7N^6+18N^5+18N^4-3N^3-21N^2-16N-4~. \N\\ \N
%
A^{Qg}_b &=&T_RC_F\Biggl\{
            \frac{1}{\ep^2}\Biggl[ 
                           -\frac{32}{N}S_1(N)
                           +\frac{32}{N}
                           \Biggr] 
           +\frac{1}{\ep}\Biggl[
                         \frac{24S_2(N)-8S^2_1(N)}
                              {N}
 \N\\  
         &&              +16\frac{N^2+7N+2}
                                 {N(N+1)(N+2)}S_1(N)
                         -32\frac{N^2+5N+2}
                                 {N(N+1)(N+2)}
       \Biggr]
     -\frac{16}{N}S_{2,1}(N)
     +\frac{40}{3N}S_3(N)
 \N\\     
   &&-\frac{4}{N}S_1(N)S_2(N)
     -\frac{4}{3N}S^3_1(N)
     -\frac{8}{N}S_1(N)\zeta_2
     +4\frac{
            N^2
            +7N
             +2}
            {N(N+1)(N+2)}S^2_1(N)
 \N\\     
   &&+4\frac{
             N^2
             -9N
              +2}
            {N(N+1)(N+2)}S_2(N)
     +\frac{8}{N}\zeta_2
     -16\frac{N^3+9N^2+8N+4}
             {N^2(N+2)^2}S_1(N)
 \N
\\     
   &&+32\frac{N^5+10N^4+30N^3+37N^2+18N+4}
           {N(N+1)^3(N+2)^2}
           \Biggl\}~.       \label{resB} \\ \N 
\\
A^{Qg}_c&=&T_RC_F\Biggr\{
      -\frac{1}{\ep^2}\frac{8}{N}
     +\frac{1}{\ep}\frac{4(13N^4+82N^3+82N^2+N-6)}{N^2(N+1)(N+2)(N+3)}
     +\frac{20}{N}S_2(N)
     -\frac{2}{N}\zeta_2
 \N
\\
     &&-\frac{2P_2(N)}
            {N^3(N+1)^2(N+2)^2(N+3)}
       \Biggl\}~,       \label{resC}
\end{eqnarray} \begin{eqnarray}
P_2(N)&=&16N^7+176N^6+520N^5+600N^4+257N^3+7N^2+16N+12~.  \N\\ \N\\
%
 A^{Qg}_d&=&T_RC_F\Biggr\{
       -\frac{1}{\ep^2}\frac{16}{N}
       +\frac{1}{\ep}\Biggl[
        -\frac{8}{N}S_1(N)
        +8\frac{
               N^3
               +10N^2
               +59N
               +42}
              {N(N+1)(N+2)(N+3)}
       \Biggr]
       -\frac{2}{N}\Bigl[S_2(N)+S^2_1(N)\Bigr]
 \N\\
     &&-\frac{4}{N}\zeta_2
       +4\frac{N^4+8N^3+43N^2+36N+12}
              {N^2(N+1)^2(N+2)}S_1(N)  \N
\\
     &&-\frac{8P_3(N)}{N(N+1)^3(N+2)^2(N+3)}
         \Biggl\}~,       \label{resD} \\
P_3(N)&=&N^6+10N^5+99N^4+350N^3+486N^2+274N+60~. \N\\ \N\\
%
 A^{Qg}_e&=&T_R\Biggl[C_F-\frac{C_A}{2}\Biggr] \Biggl\{
       \frac{1}{\ep^2}\frac{16(N+3)}{(N+1)^2}
      +\frac{1}{\ep}\Biggl[
        -\frac{8(N+2)}{N(N+1)}S_1(N) \N\\
      &&-8\frac{
               3N^3
               +9N^2
               +12N
               +4}
               {N(N+1)^3(N+2)}
      \Biggr]
      -2\frac{9N^4+40N^3+71N^2-12N-36}
             {N(N+1)^2(N+2)(N+3)}S_2(N) \N\\
    &&-2\frac{N^3-N^2-8N-36}
             {N(N+1)(N+2)(N+3)}S^2_1(N)
      +4\frac{(N+3)}{(N+1)^2}\zeta_2 \N\\
    &&+4\frac{4N^5+19N^4+31N^3-30N^2-44N-24}
             {N^2(N+1)^2(N+2)(N+3)}S_1(N)\N\\
    &&+\frac{4P_4(N)}
             {N^2(N+1)^4(N+2)^2(N+3)}   \Biggr\}~,   \label{resE} \\
P_4(N)&=&16N^7+111N^6+342N^5+561N^4+536N^3+354N^2+152N+24~. \N \\\N\\
%
A^{Qg}_f&=&T_R\Biggl[C_F-\frac{C_A}{2}\Biggr]\Biggl\{
           \frac{1}{\ep^2}\Biggl[
                          \frac{64}{(N+1)(N+2)}S_1(N)
                         -\frac{64}{(N+1)(N+2)}
                         \Biggr] \N\\
        &&   +\frac{1}{\ep}\Biggl[
                         -\frac{16}{N}S_2(N)
                         +16\frac{5N+2}
                                 {N^2(N+1)(N+2)}S_1(N)
                         -\frac{32}{(N+1)^2(N+2)}
                        \Biggr] \N\\
                  &&     +\frac{16}{N}S_{2,1}(N)
                         -\frac{8}{N}S_3(N)
                         +\frac{16}{(N+1)(N+2)}S_1(N)\zeta_2 \N\\
                  &&     +4\frac{(9N+2)(2N-3)}
                                {N^2(N+1)(N+2)}S_2(N) 
                         +4\frac{2N^2-3N+2}
                                {N^2(N+1)(N+2)}S^2_1(N)\N\\
                  &&     -\frac{16}{(N+1)(N+2)}\zeta_2
                         -8\frac{17N^2+32N+12}
                         {N(N+1)^2(N+2)^2}S_1(N) \N\\
                  &&     +16\frac{2N^3+12N^2+23N+18}
                         {(N+1)^3(N+2)^2}  
                       \Biggr\}~.
      \label{resF} \\\N\\
 A^{Qg}_g&=&T_RC_F\Biggl\{
       \frac{1}{\ep^2}\frac{32}{(N+1)(N+2)}
       +\frac{1}{\ep}\Biggl[
         \frac{8}{(N+1)(N+2)}S_1(N) \N\\
       &&-8\frac{
                17N^2
                +47N
                +28}
                {(N+1)^2(N+2)^2}
       \Biggr] 
%
%
         -\frac{38}
               {(N+1)(N+2)}S_2(N) 
       +\frac{2}{(N+1)(N+2)}S^2_1(N) \N
\end{eqnarray}\begin{eqnarray}
     &&+\frac{8}
              {(N+1)(N+2)}\zeta_2
         -4\frac{3N^3+31N^2+45N+8}
               {N(N+1)^2(N+2)^2}S_1(N) \N\\
     &&+36\frac{4N^4+26N^3+55N^2+43N+8}
                {(N+1)^{3}(N+2)^{3}} \Biggr\}~.      \label{resG}
\\
%
 A^{Qg}_h&=&T_R\Biggl[C_F-\frac{C_A}{2}\Biggr]\Biggl\{
       -\frac{1}{\ep^2}\frac{32}{(N+1)(N+2)}
       +\frac{1}{\ep}\Biggl[
          16\frac{N+3}
                 {N(N+1)(N+2)}S_1(N) \N\\
     &&   -8\frac{N^2+7N+8}
                 {(N+1)^2(N+2)^2}
       \Biggr]
      -4\frac{N^2-18N+9}
             {N(N+1)(N+2)(N+3)}S_2(N)
 \N\\
    &&-4\frac{N^2-2N+9}
             {N(N+1)(N+2)(N+3)}S^2_1(N)
      -\frac{8}{(N+1)(N+2)}\zeta_2
 \N\\
    &&+4\frac{3N^4+N^3-27N^2-85N-84}
              {N(N+1)^2(N+2)^2(N+3)}S_1(N) \N\\
    &&-4\frac{(N-1)(14N^4+110N^3+337N^2+463N+240)}
              {(N+1)^3(N+2)^{3}(N+3)}  \Biggr\}~. \label{resH} \\  \N\\
A^{Qg}_i&=&
T_RC_A \Biggl\{ 
    \frac{1}{\ep^2}\Biggl[
                   \frac{16}{(N+1)(N+2)}S_1(N)
                  -\frac{16(N+4)}{(N+1)(N+2)^2}
                   \Biggr]
   +\frac{1}{\ep}\Biggl[
                  \frac{32}{(N+2)}S_{-2}(N) \N\\
             &&  +\frac{4(4N+3)}{(N+1)(N+2)}S_2(N)
                 -\frac{4}{(N+1)(N+2)}S^2_1(N) \N\\
             &&  +8\frac{N^3+9N^2+17N+8}{N(N+1)^2(N+2)^2}S_1(N) 
              -8\frac{2N^3+8N^2+19N+16}{(N+1)^2(N+2)^3}
                 \Biggr]
               -\frac{32}{N+2}S_{-2,1}(N) \N\\
             &&-\frac{8(2N+1)}{(N+1)(N+2)}S_{2,1}(N)
               +\frac{16}{N+2}S_{-3}(N)
               +\frac{4(18N+17)}{3(N+1)(N+2)}S_3(N) \N\\
             &&+\frac{32}{N+2}S_{-2}(N)S_1(N)
               +\frac{2(8N+7)}{(N+1)(N+2)}S_{2}(N)S_1(N)
               -\frac{2}{3(N+1)(N+2)}S_1^3(N) \N\\
             &&+\frac{4}{(N+1)(N+2)}\zeta_2S_1(N) 
              -\frac{16(N^2-N-4)}{(N+1)(N+2)^2}S_{-2}(N) \N\\
            &&-2\frac{4N^4+N^3-7N^2+7N+8}{N(N+1)^2(N+2)^2}S_2(N)
               +2\frac{3N^3+7N^2-3N-8}{N(N+1)^2(N+2)^2}S_1^2(N)\N\\
            && -\frac{4(N+4)}{(N+1)(N+2)^2}\zeta_2
              -4\frac{4N^5+36N^4+114N^3+174N^2+137N+48}{N(N+1)^3(N+2)^3}S_1(N) 
\N\\
            &&+4\frac{8N^5+68N^4+247N^3+449N^2+403N+144}{(N+1)^3(N+2)^4}
     \Biggr\}\N\\
&&+T_RC_F \Biggl\{
        \frac{1}{\ep}\Biggl[
                    -\frac{16}{(N+1)(N+2)}S_2(N)
                    +\frac{16}{(N+1)(N+2)}S_1^2(N)\N\\
               &&   -\frac{64}{N(N+2)}S_1(N)
                 +\frac{128}{(N+1)(N+2)}
                    \Biggr]
                -\frac{32}{3(N+1)(N+2)}S_3(N)\N\\
              &&+\frac{8}{(N+1)(N+2)}S_2(N)S_1(N)
                +\frac{8}{3(N+1)(N+2)}S^3_1(N) \N
\end{eqnarray} \begin{eqnarray}
              &&+\frac{8(3N+2)}{N(N+1)(N+2)}S_2(N) 
                -\frac{8(3N-2)}{N(N+1)(N+2)}S_1^2(N)\N\\
              &&+\frac{16(5N^2+9N+6)}{N(N+1)^2(N+2)}S_1(N)
                -\frac{192}{(N+1)(N+2)}
      \Biggr\}.
      \label{resI}
\\
%
 A^{Qg}_j &=&
         -2 T_R C_A
         \exp\Biggl(\sum_{i=2}^{\infty}\frac{\zeta_i}{i}\ep^i\Biggr)
         \frac{\Gamma(N-\ep/2)\Gamma(N)}
         {\Gamma(N+2+\ep/2)\Gamma(N+3-\ep)}
         \frac{B(1-\ep/2,\ep/2)}{\ep(\ep+2)} \N\\
 &&      \Bigl(
                     4(N+2)(4N^2+4N-5)
                     -4(11N^2+9N+9)\ep
                     -(4N^3-2N^2-27N-2)\ep^2 \N\\
&&                   +(4N^2+2N+9)\ep^3
                     -2(2N-1)\ep^4
         \Bigr) \N\\
&=&T_RC_A\Biggl\{
       -\frac{1}{\ep^2}\frac{8(4N^2+4N-5)}
               {N^2(N+1)^2}
       +\frac{1}{\ep}\frac{4(4N^5+22N^4+11N^3+13N^2+35N+10)}
              {N^3(N+1)^3(N+2)}
 \N\\
    &&   -4\frac{4N^2+4N-5}{N^2(N+1)^2}S_2(N)
         -2\frac{4N^2+4N-5}{N^2(N+1)^2}\zeta_2
         -\frac{2P_5(N)}
              {N^4(N+1)^4(N+2)^2}   \Biggr\} + O(\varepsilon) ~,    \label{resJ} \\
P_5(N)&=&20N^7+64N^6+120N^5+94N^4-140N^3-253N^2-100N-20~.  \N\\\N\\
%
A^{Qg}_k&=&
         4 T_R C_A \exp\Biggl(\sum_{i=2}^{\infty}\frac{\zeta_i}{i}\ep^i\Biggr)
         \frac{\Gamma(N+1-\ep/2)\Gamma(N-1)}
         {\Gamma(N+2+\ep/2)\Gamma(N+3-\ep)}
         \frac{B(1-\ep/2,\ep/2)}{\ep(\ep+2)}
         \Bigl(
               2(3N^2-23N-20) \N\\
 &&            -(7N^2+9N+36)\ep
              +2(N^2+4N+1)\ep^2
              +(4N+9)\ep^3
              +2\ep^4
          \Bigr) \N\\
&=&T_RC_A\Biggl\{
      \frac{1}{\ep^2}\frac{8(3N^2-23N-20)}
             {(N-1)N(N+1)^2(N+2)}
      -\frac{1}{\ep}\frac{4(10N^4+7N^3+51N^2+172N+112)}
                         {(N-1)N(N+1)^3(N+2)^2}
 \N\\
      &&+4\frac{3N^2-23N-20}
             {(N-1)N(N+1)^2(N+2)}S_2(N)
        +2\frac{3N^2-23N-20}
             {(N-1)N(N+1)^2(N+2)}\zeta_2 \N\\
       &&+\frac{2P_6(N)}
            {(N-1)N(N+1)^4(N+2)^3}  \Biggr\} + O(\varepsilon)~,     \label{resK} \\
P_6(N)&=&14N^6+56N^5+153N^4+139N^3-414N^2-908N-448~.\N\\\N\\
%
A^{Qg}_l&=&T_RC_A\Biggl\{
              \frac{1}{\ep^2}\Biggl[
                             \frac{16}{N}S_1(N)
                            +8\frac{2N^3+5N^2+4N+2}
                                   {N^2(N+1)^2}
                              \Biggr]
            +\frac{1}{\ep}\Biggl[
                          \frac{4}{N}S_{2}(N) 
                          +\frac{4}{N}S^2_{1}(N) \N\\
                        &&-\frac{16}{N(N+1)}S_1(N)
                          -4\frac{4N^6+30N^5+55N^4+38N^3+4N^2-10N-4}
                                 {N^3(N+1)^3(N+2)}
                         \Biggr] \N\\
           &&  +\frac{8}{N}S_{2,1}(N)
               +\frac{4}{3N}S_{3}(N) 
               +\frac{2}{N}S_2(N)S_{1}(N) 
               +\frac{2}{3N}S^3_{1}(N) 
               +\frac{4}{N}S_1(N)\zeta_2 \N\\
           &&  -4\frac{2N^3+2N^2-N-2}
                      {N^2(N+1)^2}S_2(N)
               -\frac{4}{N(N+1)}S^2_{1}(N)
               +2\frac{2N^3+5N^2+4N+2}
                     {N^2(N+1)^2}\zeta_2 \N\\
          &&   -4\frac{(N+2)(2N+1)}
                     {N^2(N+1)^2}S_1(N)
               +2\frac{P_7(N)}
                     {N^4(N+1)^4(N+2)}
                     \Biggr\}~,
                     \label{resL} 
\end{eqnarray}\begin{eqnarray}
P_7(N)&=&8N^8+68N^7+164N^6+171N^5+78N^4+12N^3+14N^2+14N+4~. \N\\\N\\
A^{Qg}_m&=&T_RC_A\Biggl\{
                \frac{1}{\ep^2}\frac{8(N^2-2N-2)}
                                    {N^2(N+1)^2}
               -\frac{1}{\ep}\frac{4(2N^5+11N^4+12N^3+2N^2+6N+4)}
                                  {N^3(N+1)^3(N+2)}\N\\
            && +4\frac{N^2-2N-2}{N^2(N+1)^2}S_2(N)
               +2\frac{N^2-2N-2}{N^2(N+1)^2}\zeta_2
               +\frac{2P_8(N)}
                      {N^4(N+1)^4(N+2)}
                 \Biggr\}~,
      \label{resM} \\
P_8(N)&=&2N^6+7N^5+12N^4+6N^3-8N^2-10N-4~.\N \\\N\\
A^{Qg}_n&=&T_RC_A\Biggl\{
       \frac{1}{\ep^2}\Biggl[
                       8\frac{2N^2+3N+2}
                             {N(N+1)(N+2)}S_1(N)
                      -8\frac{N(N+3)}
                             {(N+1)^2(N+2)}
                      \Biggr]   \N\\
     &&+\frac{1}{\ep}\Biggl[
                          -16\frac{N-1}
                                 {N(N+1)}S_{-2}(N)
                          -2\frac{10N^2+21N+6}
                                {N(N+1)(N+2)}S_2(N)\N\\
                       && +2\frac{2N^2+3N+2}
                                {N(N+1)(N+2)}S^2_1(N)
                          -4\frac{N^5+6N^4+4N^3-30N^2-40N-8}
                                 {N^2(N+1)^2(N+2)^2}S_1(N) \N
\\
                       && +4\frac{2N^4+11N^3+15N^2+12N+8} 
                                {(N+1)^3(N+2)^2}
                    \Biggr]
             +16\frac{N-1}
                    {N(N+1)}S_{-2,1}(N) \N\\
           &&+4\frac{4N^2+5N-2}
                    {N(N+1)(N+2)}S_{2,1}(N) 
             -8\frac{N-1}
                    {N(N+1)}S_{-3}(N)\N\\
           &&-2\frac{28N^2+45N-14}
                   {3N(N+1)(N+2)}S_3(N)
             -16\frac{N-1}
                    {N(N+1)}S_{-2}(N)S_1(N)\N\\
           && -\frac{6N^2+5N-18}
                       {N(N+1)(N+2)}S_2(N)S_1(N)
             +\frac{2N^2+3N+2}
                   {3N(N+1)(N+2)}S^3_1(N) \N\\
           &&+2\frac{2N^2+3N+2}
                   {N(N+1)(N+2)}\zeta_2S_1(N)
             +16\frac{N^2-N-4} 
                    {(N+1)^2(N+2)}S_{-2}(N) \N\\
           &&+\frac{7N^5+26N^4+16N^3-58N^2-88N-24}
                   {N^2(N+1)^2(N+2)^2}S_2(N)\N\\
           &&-\frac{N^5+6N^4+4N^3-30N^2-40N-8}
                   {N^2(N+1)^2(N+2)^2}S^2_1(N)
             -2\frac{N(N+3)}
                   {(N+1)^2(N+2)}\zeta_2 \N\\
           &&+2\frac{P_9(N)}
                   {N(N+1)^3(N+2)^3}S_1(N)
             -2\frac{P_{10}(N)}
                   {(N+1)^4(N+2)^3}
             \Biggr\}
      \label{resN}~, \\
P_9(N)&=&2N^6+20N^5+40N^4-45N^3-170N^2-100N+8 ~,
\N\\
P_{10}(N)&=&4N^6+32N^5+91N^4+123N^3+62N^2-32N-40~.
\N\\\N\\
%
A^{Qg}_o&=&T_RC_A\Biggl\{
         \frac{1}{\ep^2}\Biggl[         
                              -\frac{16}{N(N+2)}S_1(N)
                              -8\frac{N^2+7N+8}
                                      {(N+1)^2(N+2)^2}
                        \Biggr]\N\\
      &&+\frac{1}{\ep}\Biggl[
                              -\frac{4}{N(N+2)}S_2(N)
                              -\frac{4}{N(N+2)}S_1^2(N)
                              +4\frac{2N^2+9N+12}
                                     {N(N+1)(N+2)^2}S_1(N) \N\\
                           && +4\frac{(11N^3+56N^2+92N+49)N}
                                   {(N+1)^3(N+2)^3}
                      \Biggr]
        -\frac{8}{N(N+2)}S_{2,1}(N)\N
\end{eqnarray}\begin{eqnarray}
      &&-\frac{4}{3N(N+2)}S_3(N)
        -\frac{2}{N(N+2)}S_2(N)S_1(N)
        -\frac{2}{3N(N+2)}S_1^3(N)\N\\
      &&-\frac{4}{N(N+2)}S_1(N)\zeta_2
        +\frac{10N^3+31N^2+41N+28}
               {N(N+1)^2(N+2)^2}S_2(N) \N\\
      &&+\frac{2N^2+9N+12}
               {N(N+1)(N+2)^2}S_1^2(N)
        -2\frac{N^2+7N+8}
               {(N+1)^2(N+2)^2}\zeta_2\N\\
      &&+2\frac{4N^4+16N^3-4N^2-61N-48}
               {N(N+1)^2(N+2)^3}S_1(N)
        -2\frac{P_{11}(N)}
               {(N+1)^4(N+2)^4}
        \Biggr\}~, \label{resO} \\
P_{11}(N)&=&28N^6+222N^5+684N^4+1038N^3+811N^2+321N+64~. \N\\\N\\
%
A^{Qg}_p&=&T_RC_A \Biggr\{
      \frac{1}{\ep^2}\Biggl[
        -8\frac{(N-4)}
                {N(N+1)(N+2)}S_1(N)
        -8\frac{N+4}
                {(N+1)(N+2)^2}
      \Biggr] \N\\
      &&+\frac{1}{\ep}\Biggl[
        2\frac{3N+4}
               {N(N+1)(N+2)}S_2(N)
         -2\frac{N-4}
               {N(N+1)(N+2)}S^2_1(N) \N
\\
      &&+4\frac{N^3-17N^2-41N-16}
               {N(N+1)^2(N+2)^2}S_1(N)
        +4\frac{4N^3+26N^2+51N+32}
               {(N+1)^2(N+2)^3}
      \Biggr]
\N\\
%
    &&-4\frac{N-4}
             {N(N+1)(N+2)}S_{2,1}(N)
      +\frac{2}{3}\frac{5N+4}
               {N(N+1)(N+2)}S_3(N) \N\\
    &&-\frac{1}{3}\frac{N-4}
               {N(N+1)(N+2)}S^3_1(N)
      -\frac{N-4}
             {N(N+1)(N+2)}S_1(N)S_2(N) \N\\
     && -2\frac{N-4} 
             {N(N+1)(N+2)}S_1(N)\zeta_2
       -\frac{7N^3+17N^2+13N+16}
             {N(N+1)^2(N+2)^2}S_2(N) \N\\
    &&+\frac{N^3-17N^2-41N-16}
             {N(N+1)^{2}(N+2)^{2}}S^2_1(N)
      -2\frac{N+4}
             {(N+1)(N+2)^2}\zeta_2 \N\\
    && +2\frac{2N^5+48N^4+174N^3+242N^2+161N+64}
             {N(N+1)^3(N+2)^3}S_1(N) \N\\
    &&-2\frac{10N^5+92N^4+329N^3+581N^2+507N+176}
             {(N+1)^3(N+2)^4}   \Biggl\}~.    \label{resP}\\ \N\\
A^{Qg}_q&=& A^{Qg}_r = A^{Qg}_{r'} = 0
\\
A^{Qg}_s&=&T_RC_A\Biggl\{
      -\frac{1}{\ep^2}\frac{8}{N^2(N+1)^2}
      +\frac{1}{\ep}\frac{4(2N^3+N^2-3N-1)}
                         {N^3(N+1)^3}
     -\frac{4}{N^2(N+1)^2}S_2(N)\N\\
   &&-\frac{2}{N^2(N+1)^2} \zeta_2 
     -\frac{2P_{12}(N)}
              {N^4(N+1)^4(N+2)}\Biggr\}~, \label{resS} \\
P_{12}(N)&=&4N^6+4N^5-8N^4-2N^3+16N^2+9N+2~. \N\\
%
A^{Qg}_t&=&T_RC_A\Biggl\{
   \frac{1}{\ep^2}\frac{8(N^2+3N+4)}
                        {(N-1)N(N+1)^2(N+2)}
   -\frac{1}{\ep}\frac{4(2N^4+5N^3-3N^2-20N-16)}
          {(N-1)N(N+1)^3(N+2)^2} \N\\
  &&+4\frac{N^2+3N+4}
          {(N-1)N(N+1)^2(N+2)}S_2(N)
   +2\frac{N^2+3N+4}
          {(N-1)N(N+1)^2(N+2)}\zeta_2 \N
\end{eqnarray}\begin{eqnarray}
   &&+\frac{2P_{13}(N)}{(N-1)N(N+1)^4(N+2)^3}
           \Biggr\}~, \label{resT} \\
P_{13}(N)&=&2N^6+4N^5-13N^4-35N^3+14N^2+92N+64~. \N
\end{eqnarray}
%
%
The pure-singlet contributions read~:
\begin{eqnarray}
A^{Qq}_{a}&=& T_RC_F\Biggl\{
         -\frac{1}{\ep^2}\frac {16(N^2+N-2)}
                 {N^2(N+1)^2}
         -\frac{1}{\ep}\frac{8(5N^3-5N^2-16N-4)}
                {N^3(N+1)^3(N+2)} \N\\
         &&-8\frac{N^2+N-2}
                {N^2(N+1)^2}S_2(N)
         -4\frac{N^2+N-2}
                {N^2(N+1)^2}\zeta(2)
         +\frac{4P_{14}(N)}
                {N^4(N+1)^4(N+2)^2} \Biggr\}~, \label{psresA}\\
P_{14}(N)&=&N^8+7N^7+16N^6-9N^5-26N^4+61N^3+110N^2+44N+8~.   \N\\\N\\
%
A^{Qq}_{b}&=&T_RC_F\Biggl\{
          -\frac{1}{\ep^2}\frac{128}
               {(N-1)N(N+1)(N+2)} 
         -\frac{1}{\ep}\frac{128(3+2N)}
           {(N-1)N(N+1)^2(N+2)^2} \N\\
          &&-\frac{64}
                  {(N-1)N(N+1)(N+2)}S_2(N)
          -\frac{32}
                  {(N-1)N(N+1)(N+2)}\zeta_2 \N\\
          &&+\frac{32(N^4+6N^3+N^2-24N-24)}
                  {(N-1)N(N+1)^3(N+2)^3} \Biggr\}~. \label{psresB}\N\\
%
%
\end{eqnarray}
Finally the  non-singlet contributions are~:
\begin{eqnarray}
%
%
%
  A_a^{qq,Q}&=&T_RC_F\Biggl\{
          -\frac{1}{\ep^2}\frac{8(N^2-2+N)}
                               {3N(N+1)}
          -\frac{1}{\ep}  \frac{8(N^4+2N^3-10N^2-5N+3)}
                               {9N^2(N+1)^2} \N\\
                  &&    -4\frac{11N^6+33N^5-34N^4-57N^3+5N^2+6N-9}
                               {27N^3(N+1)^3}
                        -2\frac{N^2-2+N}
                               {3N(N+1)}\zeta_2
            \Biggr\}~. \label{nsresA} \\\N\\
%
  A_b^{qq,Q}&=&T_RC_F\Biggl\{
            \frac{1}{\ep^2}\Biggl[
                                -\frac{32}{3}S_1(N)
                               +\frac{32}{3}
                           \Biggr]
           +\frac{1}{\ep}\Biggl[
                               \frac{16}{3}S_2(N)
                               -\frac{80}{9}S_1(N)
                               +\frac{32}{9}
                         \Biggr] \N
\end{eqnarray}\begin{eqnarray}
                             &&-\frac{8}{3}S_3(N)
                               -\frac{8}{3}\zeta_2S_1(N)
                               +\frac{40}{9}S_2(N)
                               +\frac{8}{3}\zeta_2
                               -\frac{224}{27}S_1(N)
                               +\frac{176}{27}
              \Biggr\}~. \label{nsresB}\\\N\\
  A_c^{qq,Q}&=&
        T_RC_F\Biggl\{-\frac{2}{\ep}-\frac{5}{6} \Biggl\}~. \label{nsresC}
         \N\\
\end{eqnarray}
\newpage
%
%
 \section{\bf\boldmath Appendix~B: Finite and Infinite Sums}
 \label{App-sums}
   \renewcommand{\theequation}{\thesection.\arabic{equation}}
   \setcounter{equation}{0}
  In the following, we list several classes of sums, which were used to
  derive the results in the present paper beyond 
  well-known results for harmonic sums. Other relations  
  can be found in \cite{Devoto:1983tc,HSUM1,HSUM2,%
  Blumlein:2004bs,Blumleinunp} and were used in the present calculation.
  Here $N,L,A$ denote arbitrary integers, $a$ is a complex number, and
  $B(a,b)$ is Euler's Beta-function. 
  \subsection{\bf\boldmath Sums involving Beta-Functions}
   \begin{eqnarray}
     \sum_{i=1}^{\infty}\frac{B(N,i)}{i}
       &=&\zeta_2-S_2(N-1) 
          ~,
       \label{2b}\\ \N\\
    \sum_{i=1}^{\infty}\frac{B(N,i)}{i+1}
       &=&1+\Bigl[S_2(N-1)-\zeta_2\Bigl] (N-1) 
          ~,
          \label{5b}\\ \N\\
     \sum_{i=1}^{\infty}\frac{B(N,i)}{i+2}
       &=&\frac{5-2N}{4}+\Bigl[\zeta_2-S_2(N-1)\Bigr]
       \frac{(N-1)(N-2)}{2} 
           ~,
          \label{6b}\\ \N
\\
     \sum_{i=1}^{\infty}\frac{B(N,i)}{i+3}
       &=&\frac{6N^2-33N+49}{36}  \N\\
         &&+
           \Bigl[S_2(N-1)-\zeta_2\Bigr]\frac{(N-1)(N-2)(N-3)}{6}
          ~,
         \label{7b}\\ \N
\\
     \sum_{i=1}^{\infty}\frac{B(N,i)}{i^2}
       &=&\zeta_3-\zeta_2S_1(N-1)+S_{1,2}(N-1)
         ~,
         \label{3b}\\ \N\\
     \sum_{i=1}^{\infty}\frac{B(N,i)}{i^3}
       &=&\frac{2}{5}\zeta_2^2+\zeta_2S_{1,1}(N-1)-S_{1,1,2}(N-1)
       -\zeta_3S_1(N-1)
         ~,
         \label{4b}\\ \N\\
      \sum_{i=1}^{\infty}\frac{B(N,i)}{N+i}
       &=&\frac{1}{N^2}
          ~,
          \label{8b}\\ \N\\
     \sum_{i=1}^{\infty}\frac{B(N,i)}{1+N+i}
       &=&\frac{N^2+N+1}{N^2(N+1)^2}
           ~,
       \label{9b}\\ \N\\
     \sum_{i=1}^{\infty}\frac{B(N,i)}{2+N+i}
       &=&\frac{N^4+4N^3+7N^2+6N+4}{N^2(N+1)^2(N+2)^2}
          ~,
         \label{10b}\\ \N\\
     \sum_{i=1}^{\infty}\frac{B(N,i)}{3+N+i}
       &=&\frac{N^6+9N^5+34N^4+69N^3+85N^2+66N+36}
       {N^2(N+1)^2(N+2)^2(N+3)^2}
          ~,
           \label{12b} \\ \N\\
     \sum_{i=1}^{\infty}\frac{B(N,i)}{4+N+i}
       &=&\frac{P(N)}{N^2(N+1)^2(N+2)^2(N+3)^2(N+4)^2}~,\N
\end{eqnarray} \begin{eqnarray}
   P(N)&=&N^8+16N^7+110N^6+424N^5+1013N^4+1576N^3 \N\\
        &&+1660N^2+1200N+576~,
           \label{18b} \\ \N\\
     \sum_{i=1}^{\infty} \frac{B(N+1,i)}{N+i}
       &=&(-1)^N\Bigl[2S_{-2}(N)+\zeta_2\Bigr]
          ~,
          \label{13b}\\
     \sum_{i=1}^{\infty} \frac{B(N+2,i)}{N+i}
       &=&(N+1)(-1)^N\Bigl[2S_{-2}(N)+\zeta_2\Bigr]-\frac{1}{N+1}
       ~,
       \label{14b}\\ \N\\
     \sum_{i=1}^{\infty} \frac{B(N+1,i)}{(N+i)^2}
       &=&(-1)^N\Bigl[\zeta_3+S_1(N)\zeta_2+2S_{1,-2}(N)
        +S_{-3}(N)\Bigr]
        ~.
       \label{15b}
\\
     \sum_{i=1}^{\infty} \frac{B(N,i)}{(N+i+1)^2}
       &=&\frac{(-1)^{N}}{N(N+1)}\Bigl[2S_{-2}(N)+\zeta_2\Bigr]
       +\frac{N-1}{N(N+1)^3}
        ~,
        \label{17b}\\ \N\\
     \sum_{i=1}^{\infty}\frac{B(N,i)}{(2+N+i)^2}
       &=&2(-1)^N\frac{2S_{-2}(N+2)+\zeta_2}{N(N+1)(N+2)}
       +\frac{N^2+N+1}{N(N+1)^2(N+2)^2}
         ~,
         \label{11b}
\\
     \sum_{i=1}^{\infty} \frac{B(N,i)}{(N+i+1)^3}
       &=&\frac{(-1)^N}{N(N+1)}\Bigl[\zeta_3+S_1(N+1)\zeta_2
       -\zeta_2+2S_{1,-2}(N+1) \N\\
       &&-2S_{-2}(N+1)+S_{-3}(N+1)\Bigr]
       ~,
        \label{16b}\\ \N
   \end{eqnarray}
  \subsection{\bf\boldmath Sums involving Beta-Functions and Harmonic Sums}
   \begin{eqnarray}
     \sum_{i=1}^{\infty}B(N,i)S_1(i)
       &=&\zeta_2-S_2(N-2)
         ~,
          \label{1c}\\ \N\\
     \sum_{i=1}^{\infty}\frac{B(N,i)S_1(i)}{i}
       &=&2\zeta_3+S_1(N-1)S_2(N-1)-\zeta_2S_1(N-1)\N\\
        &&-S_{2,1}(N-1)
          ~,
           \label{2c}\\ \N\\
     \sum_{i=1}^{\infty}\frac{B(N,i)S_1(i)}{i^2}
       &=&\frac{1}{2}\zeta_2^2+\zeta_2S_{1,1}(N-1)-2\zeta_3S_1(N-1)\N\\
       &&-S_{1,1,2}(N-1)+S_{1,3}(N-1)
       ~,
        \label{21c}\\ \N
\\
     \sum_{i=1}^{\infty}\frac{B(N,i)S_1(i)}{N+i}
        &=&\frac{\zeta_2-S_2(N-1)}{N}
        ~,
       \label{3c}\\ \N\\
     \sum_{i=1}^{\infty}\frac{B(N,i)S_1(i)}{N+i+1}
       &=&\frac{\zeta_2-S_2(N-1)}{N+1}+\frac{1}{N^3(N+1)}
      ~,
        \label{4c}\\ \N\\
     \sum_{i=1}^{\infty}\frac{B(N,i)S_1(i)}{N+i+2}
       &=&\frac{\zeta_2-S_2(N+1)}{N+2}
          +\frac{2N^3+2N^2+3N+1}{N^3(N+1)^3}
         ~,
          \label{20c}\\ \N
\end{eqnarray} \begin{eqnarray}
     \sum_{i=1}^{\infty}\frac{B(N+1,i)S_1(i)}{N+i}
       &=&\frac{\zeta_2-S_2(N)}{N}
       +(-1)^N\Bigl[\zeta_3+S_{-3}(N)-2\frac{S_{-2}(N)}{N}\N\\
       &&+2S_{1,-2}(N)
       -\frac{\zeta_2}{N}+\zeta_2S_1(N)\Bigr]
       ~,
        \label{29c}\\ \N\\
     \sum_{i=1}^{\infty} B(N,i)S_1(N+i)
       &=&\frac{S_1(N-1)}{N-1}+\frac{2N^2-2N+1}{N^2(N-1)^2}
       ~,
        \label{5c}%
\\ \N\\
     \sum_{i=1}^{\infty}\frac{B(N,i)S_1(N+i)}{i}
       &=&2\zeta_3-2S_3(N)+S_1(N)\Bigl[\zeta_2-S_2(N)\Bigr]
       +\frac{S_1(N)}{N^2}\N\\
       &&+\frac{1}{N^3}
        ~,
         \label{6c}\\ \N\\
     \sum_{i=1}^{\infty}\frac{B(N,i)S_1(N+i-1)}{i}
       &=&2\zeta_3-2S_3(N-1)+S_1(N-1)\Bigl[\zeta_2-S_2(N-1)\Bigr]
        ~,
         \label{7c}\\ \N\\
     \sum_{i=1}^{\infty} \frac{B(N,i)S_1(N+i)}{N+i}
       &=&\frac{S_1(N-1)}{N^2}+\frac{2}{N^3}
         ~,
          \label{8c}\\ \N\\
     \sum_{i=1}^{\infty} \frac{B(N,i)S_1(N+i)}{N+1+i}
       &=&\frac{N^2+N+1}{N^2(N+1)^2}S_1(N)
       -\frac{(-1)^N}{N(N+1)}\Bigl[2S_{-2}(N)+\zeta_2\Bigr] \N\\
       &&+\frac{1}{N^3}
         ~,
          \label{10c} \\ \N\\ 
     \sum_{i=1}^{\infty} \frac{B(N,i)S_1(N+i-1)}{N+1+i}
       &=&\frac{N^2+N+1}{N^2(N+1)^2}S_1(N)
       -\frac{(-1)^N}{N(N+1)}\Bigl[2S_{-2}(N)
       +\zeta_2\Bigr] \N\\
       &&+\frac{2N+1}{N^3(N+1)^2}
        ~,
         \label{11c}\\ \N
\\
     \sum_{i=1}^{\infty} \frac{B(N,i)S_1(N+i)}{N+2+i}
       &=&\frac{N^3+2N^2+5N+2}{N^3(N+1)^2(2+N)}\N\\
        &&+\frac{N^4+4N^3+7N^2+6N+4}{N^2(N+1)^2(N+2)^2}S_1(N) \N\\
        &&-\frac{2(-1)^N}{N(N+1)(N+2)}\Bigl[2S_{-2}(N)
          +\zeta_2\Bigr]
         ~,
          \label{32c}\\ \N\\
     \sum_{i=1}^{\infty} \frac{B(N,i)S_1(N+i-1)}{(N+1+i)^2}
       &=&\frac{(-1)^N}{N(N+1)}\Bigl[
       2S_{-2,1}(N+1)-2S_{1,-2}(N+1) \N\\
       &&+4S_{-2}(N+1)
       +2\zeta_2+\zeta_3-\zeta_2S_1(N+1)\Bigr] \N\\
       &&+\frac{S_1(N+1)}{N(N+1)^2}
       +\frac{1}{N(N+1)^3}
        ~,
         \label{12c}\\ \N\\
     \sum_{i=1}^{\infty} \frac{B(N+1,i)S_1(N+i)}{N+i}
       &=&(-1)^N\Bigl[2S_{-2,1}(N)+S_{-3}(N)+2\zeta_3\Bigr]
      ~,
        \label{9c}\\ \N
\end{eqnarray} \begin{eqnarray} 
     \sum_{i=1}^{\infty}\frac{B(N,i)S_1(N+i)}{i^2}
       &=&\frac{1}{2}\zeta_2^2+\zeta_3\Bigl[\frac{2}{N}-S_1(N)\Bigr]
         +\zeta_2\Bigr[\frac{S_1(N)}{N}-2S_{1,1}(N)\Bigr] \N\\
       &&+S_{2,2}(N)+2S_{1,3}(N)+S_1(N)S_{1,2}(N)-2\frac{S_3(N)}{N}\N\\
       &&-\frac{S_1(N)S_2(N)}{N}
         ~,
          \label{34c}%
\\ \N\\
     \sum_{i=1}^{\infty}\frac{B(N,i)S_1(N+i-1)}{i^2}
       &=&\frac{1}{2}\zeta_2^2-\zeta_3S_1(N-1)-2\zeta_2S_{1,1}(N-1)
         +S_{2,2}(N-1)\N\\
       &&+2S_{1,3}(N-1)+S_1(N-1)S_{1,2}(N-1)
       ~,
        \label{33c} \\ \N\\
     \sum_{i=1}^{\infty} B(N,i)S_1(i)^2
       &=&3\zeta_3-\zeta_2S_1(N-2)+S_{1,2}(N-2)-2S_3(N-2)
         ~,
          \label{28c}\\ \N\\
     \sum_{i=1}^{\infty} B(N,i)S_2(i)
       &=&-S_1(N-2)\zeta_2+\zeta_3+S_{1,2}(N-2)
         ~,
          \label{24c}\\ \N\\
     \sum_{i=1}^{\infty} \frac{B(N,i)S_1^2(i)}{i}
       &=&+\frac{17}{10}\zeta_2^2-3\zeta_3S_1(N-1)
          +\frac{1}{2}\zeta_2\Bigl[S_1^2(N-1)\N\\
         &&-S_2(N-1)\Bigr]
           -S_{1,1,2}(N-1)+S_{2,2}(N-1)\N\\
         &&+2S_1(N-1)S_3(N-1)-2S_{3,1}(N-1)
           ~,
            \label{35c} \\ 
     \sum_{i=1}^{\infty} \frac{B(N,i)S_2(i)}{i}
       &=&S_{2,2}(N-1)-S_{1,1,2}(N-1)+\zeta_2S_{1,1}(N-1)\N\\
        &&-\zeta_2S_2(N-1)
          -\zeta_3S_1(N-1)+\frac{7}{10}\zeta_2^2
          ~,
           \label{27c}\\ \N\\
    \sum_{i=1}^{\infty} \frac{B(N,i)S_2(i)}{N+i}
       &=&\frac{-S_1(N-1)\zeta_2+\zeta_3+S_{1,2}(N-1)}{N}
         ~,
           \label{25c}\\ \N\\
     \sum_{i=1}^{\infty} \frac{B(N,i)S_2(i)}{1+N+i}
       &=&\frac{-S_1(N)\zeta_2+\zeta_3+S_{1,2}(N)}{N+1}
         -\frac{S_2(N)-\zeta_2}{N^2}
           ~,
            \label{26c}\\ \N
     \sum_{i=1}^{\infty} B(N,i)S_{1,1}(N+i)&=&
       \frac{S_{1,1}(N-1)}{N-1}+\frac{S_1(N-1)}{(N-1)^2}+\frac{1}{(N-1)^3}
       \N\\ &&+\frac{S_1(N-1)}{N^2}+\frac{2}{N^3}
       ~,
       \label{14c}\\ \N\\
     \sum_{i=1}^{\infty} B(N,i)S_{1,1}(N+i-1)&=&
       \frac{S_{1,1}(N-1)}{N-1}+\frac{S_1(N-1)}{(N-1)^2}+\frac{1}{(N-1)^3}
        ~,
       \label{13c}\\ \N\\
     \sum_{i=1}^{\infty} \frac{B(N,i)S_{1,1}(N+i)}{i}
       &=&S_{1,1}(N)\Bigl[\zeta_2-S_2(N)\Bigr]
       +\frac{6}{5}\zeta_2^2-3S_4(N) \N\\
       &&+2S_1(N)\Bigl[\zeta_3-S_3(N)\Bigr]
       +\frac{S_1(N)}{N^3}+\frac{S_{1,1}(N)}{N^2}+\frac{1}{N^4}
        ~,
         \label{19c}\\ \N
\end{eqnarray} \begin{eqnarray}
     \sum_{i=1}^{\infty} \frac{B(N,i)S_{1,1}(N+i-1)}{i}
       &=&S_{1,1}(N-1)\Bigl[\zeta_2-S_2(N-1)\Bigr]
       +\frac{6}{5}\zeta_2^2-3S_4(N-1) \N\\
       &&+2S_1(N-1)\Bigl[\zeta_3-S_3(N-1)\Bigr]
      ~,
       \label{18c}%
\\ \N\\
     \sum_{i=1}^{\infty} \frac{B(N,i)S_{1,1}(N+i)}{N+i}&=&
       \frac{1}{N}\Biggl[\frac{S_{1,1}(N)}{N}+\frac{S_1(N)}{N^2}
       +\frac{1}{N^3}\Biggr]
      ~,
       \label{15c}\\ \N\\
     \sum_{i=1}^{\infty} \frac{B(N,i)S_{1,1}(N+i)}{N+i+1}&=&
       \frac{S_1(N)}{N^3}+\frac{S_{1,1}(N)(N^2+N+1)}{N^2(N+1)^2}
       +\frac{1}{N^4}  \N\\
       &&-\frac{(-1)^N}{N(N+1)}\Bigl[
       2\zeta_3+S_{-3}(N)+2S_{-2,1}(N)\Bigr]
        ~,
       \label{17c} \\ \N\\
     \sum_{i=1}^{\infty} \frac{B(N,i)S_{1,1}(N+i-1)}{N+i+1}&=&
       \frac{S_1(N-1)}{N^2(N+1)}+\frac{S_{1,1}(N-1)(N^2+N+1)}{N^2(N+1)^2}\N\\
       &&-\frac{(-1)^N}{N(N+1)}\Bigl[\zeta_2
       +2\zeta_3+2S_{-2}(N-1) \N\\ 
       &&+S_{-3}(N-1)
         +2S_{-2,1}(N-1)\Bigr]
       ~,
       \label{16c} \\ \N
     \sum_{i=1}^{\infty} B(N,i)S_1(i)S_{1}(N+i)
       &=&S_1(N-1)\Bigl[\zeta_2-S_2(N-1)\Bigr]
          +2\Bigl[\zeta_3-S_3(N-1)\Bigr]\N\\
       &&   +\frac{\zeta_2}{N}
         -\frac{S_2(N-1)}{N}+\frac{S_1(N-1)}{(N-1)^2}+\frac{2}{(N-1)^3}
        ~,
         \label{22c}\\ \N\\
     \sum_{i=1}^{\infty} \frac{B(N,i)S_1(i)S_{1}(N+i)}{i}
       &=&\frac{17}{10}\zeta_2^2+2\frac{\zeta_3}{N}-\zeta_2\Bigl[
          \frac{1}{N^2}+\frac{S_1(N-1)}{N}+2S_{1,1}(N-1)\Bigr] \N\\
          &&+\frac{S_2(N-1)}{N^2}-\frac{S_3(N-1)}{N}+\frac{S_{1,2}(N-1)}{N}\N\\
          &&+S_1(N-1)\Bigl(S_3(N-1)
            +S_{1,2}(N-1)\Bigr)\N\\
          &&+S_2(N-1)^2-S_{2,2}(N-1)-2S_{3,1}(N-1)
         ~,
          \label{30c}\\ \N\\ 
     \sum_{i=1}^{\infty} \frac{B(N,i)S_1(i)S_{1}(N+i)}{N+i}
       &=&\frac{1}{N}\Bigl\{
          S_1(N)\Bigl[\zeta_2-S_2(N)\Bigr]
          +2\Bigl[\zeta_3-S_3(N)\Bigr] \N\\
        &&+\frac{S_1(N)}{N^2}
          +\frac{2}{N^3}\Bigr\}
         ~,
          \label{23c}\\ \N\\
     \sum_{i=1}^{\infty} \frac{B(N,i)S_1(i)S_{1}(N+i)}{1+N+i}
       &=&\frac{(-1)^{N+1}}{N(N+1)}\Biggl[
          \zeta_3+\zeta_2S_1(N)+S_{-3}(N)+2S_{1,-2}(N)\Biggr]\N\\
          &&+\frac{\zeta_2S_1(N)}{N+1}+S_1(N)\Biggl[\frac{1}{N^3}-
          \frac{S_2(N)}{N+1}\Biggr]
          +\frac{2\zeta_3}{N+1}\N\\
          &&-\frac{2S_3(N)}{N+1}+\frac{2}{N^4}
          ~,
          \label{31c} \\ \N
   \end{eqnarray} \begin{eqnarray}
     \sum_{k=1}^{\infty}\sum_{i=1}^{\infty}\frac{
\Gamma(N+k)}{\Gamma(N) \Gamma(k+1)}
       \frac{B(k+i,N)}{k}\frac{S_1(i)}{i}
       &=&\frac{17}{10}\zeta_2^2-2\zeta_3S_1(N-1)+S_{2,2}(N-1)\N\\
       &&+S_1(N-1)S_{2,1}(N-1)
       -2S_{2,1,1}(N-1)
        ~.
         \label{h4}
    \end{eqnarray}
  \subsection{\bf\boldmath Sums involving Beta-Functions and Harmonic Sums 
                           with two Free Indexes}
   \begin{eqnarray}   
     \sum_{i=1}^{\infty}B(N-a,i+a)
       &=&B(1+a,N-1-a)
          ~,
       \label{1b}\\
     \sum_{i=1}^{\infty} \frac{B(N,i+k)}{i}
       &=&-\frac{d}{dN}B(k,N)\N\\
       &=&B(k,N)\Bigl[S_1(k+N-1)-S_1(N-1)\Bigr]
         ~,\label{h1} \\
     \sum_{i=1}^{\infty} \frac{B(N,i+k)}{i}S_1(i+N+k-1)
       &=&B(k,N)\Bigl[2S_{1,1}(k+N-1)-S_2(N-1)  \N\\
       &&-S_1(N+k-1)S_1(N-1)\Bigr]
        ~,\label{h2} 
\end{eqnarray} \begin{eqnarray}
     \sum_{i=1}^{\infty} \frac{B(N,i+k)}{i}S_1(i+k-1)
       &=&B(k,N)\Bigl[S_{1}(k+N-1)S_1(k-1)
         -S_2(N-1) \N\\
       &&-S_1(k-1)S_1(N-1) +\zeta_2\Bigr]
       ~.\label{h3}
   \end{eqnarray}   
  \subsection{\bf\boldmath Sums involving Binomials}
   \begin{eqnarray}
     \sum_{k=0}^{N-1}\binom{N-1}{k}(-1)^k\frac{1}{(k+A)^3}
       &=&\frac{B(N,A)}{2}\Biggl[
           \Bigl\{S_1(N+A-1)-S_1(A-1)\Bigr\}^2 \N\\
        && -S_2(A-1)+S_2(N+A-1)\Biggr]
           ~,\label{13d}\\ \N\\ 
     \sum_{k=1}^{N-1}\binom{N-1}{k}(-1)^k\frac{1}{k}
       &=&\lim_{\ep \rightarrow 0} \Bigl\{
        B(N,\ep)-\frac{1}{\ep}  \Bigr\}
       =-S_1(N-1)
       ~,\label{18e}\\ \N\\ 
     \sum_{k=1}^{N-1}\binom{N-1}{k}(-1)^k\frac{1}{k^2}
       &=&-S_{1,1}(N-1)
       ~,\label{24e}\\ \N\\ 
     \sum_{k=1}^{N-1}\binom{N-1}{k}(-1)^k\frac{1}{k^3}
       &=&-S_{1,1,1}(N-1)
       ~,\label{25e}\\ \N\\ 
     \sum_{k=1}^{N-1}\binom{N-1}{k}(-1)^k\frac{1}{(k+1)^3}
       &=&\frac{S_{1,1}(N)}{N}-1
       ~,\label{5d}\\ \N
\end{eqnarray}\begin{eqnarray}
     \sum_{k=1}^{N-1}\binom{N-1}{k}(-1)^k\frac{1}{(k+2)^3}
       &=&\frac{S_{1,1}(N)}{N(N+1)}-\frac{S_1(N)}{(N+1)^2}-\frac{1}{(N+1)^3}
       -\frac{1}{8}
       ~,\label{6d} \\ \N
     \sum_{k=1}^{N-1}\binom{N-1}{k}(-1)^k\frac{1}{(k+3)^3}
       &=&\frac{2S_{1,1}(N)}{N(N+1)(N+2)}-\frac{(5+3N)S_1(N)}{(N+1)^2(N+2)^2}
       \N\\
       &&+\frac{N^3-8N-9}{(N+1)^3(N+2)^3}
         -\frac{1}{27}
       ~,\label{7d}\\ \N
\\
     \sum_{k=1}^{N-1}\binom{N-1}{k}(-1)^k\frac{1}{(k+1)^4}
       &=&\frac{S_{1,1,1}(N)}{N}-1 
       ~,\label{8d}\\ \N\\    
     \sum_{k=1}^{N-1}\binom{N-1}{k}(-1)^k\frac{1}{(k+2)^4}
       &=&\frac{S_{1,1,1}(N)}{N(N+1)}-\frac{S_{1,1}(N)}{(N+1)^2}
       -\frac{S_{1}(N)}{(N+1)^3}-\frac{1}{(N+1)^4}\N\\
       &&-\frac{1}{16}
       ~,\label{9d}\\ \N\\
     \sum_{k=1}^{N-1}\binom{N-1}{k}(-1)^k\frac{1}{(k+3)^4}
       &=&\frac{2S_{1,1,1}(N)}{N(N+1)(N+2)}
       -\frac{(5+3N)S_{1,1}(N)}{(N+1)^2(N+2)^2} \N\\
       &&+\frac{(N^3-8N-9)S_{1}(N)}{(N+1)^3(N+2)^3} \N\\
       &&+\frac{-17-21N-2N^2+6N^3+2N^4}{(N+1)^4(N+2)^4}
       -\frac{1}{81}
       ~.\label{10d}\\ \N
   \end{eqnarray}
  \subsection{\bf\boldmath Sums involving Binomials and Harmonic Sums}
   We set  $S_A(0):=0$. Therefore the sums below may begin at $k=0$.
   \begin{eqnarray} 
     \sum_{k=0}^{N-1}\binom{N-1}{k}(-1)^k\frac{S_1(k)}{k}
       &=&\lim_{\ep \rightarrow 0} \Bigl\{
        B(N,\ep)\Bigl[S_1(N+\ep-1)-S_1(N-1)\Bigr]
        \Bigr\}  \N\\ 
       &=&\zeta_2-S_2(N-1)
       ~,\label{26e}%
\\ \N\\ 
     \sum_{k=1}^{N-1}\binom{N-1}{k}(-1)^k\frac{S_1(k)}{k^2}
       &=&-S_{1,2}(N-1)
       ~,\label{19e}\\ \N\\ 
     \sum_{k=0}^{N-1}\binom{N-1}{k}(-1)^k\frac{S_1(k+1)}{(k+1)^2}
       &=&\frac{S_2(N)}{N}
       ~,\label{13e}\\ \N\\ 
     \sum_{k=0}^{N-1}\binom{N-1}{k}(-1)^k\frac{S_1(k+2)}{(k+2)^2}
       &=&\frac{S_2(N)}{N(N+1)}-\frac{1}{(N+1)^3}
       ~,\label{14e}\\ \N%
\\   
     \sum_{k=0}^{N-1}\binom{N-1}{k}(-1)^k\frac{S^2_1(k+1)}{k+1}
       &=&-\frac{S_1(N)}{N^2}+\frac{2}{N^3}
       ~,\label{15e}\\ \N
\end{eqnarray}\begin{eqnarray}
     \sum_{k=0}^{N-1}\binom{N-1}{k}(-1)^k\frac{S^2_1(k+2)}{k+2}
       &=&-\frac{2N+1}{N^2(N+1)^2}S_1(N)
          +\frac{N^3+6N^2+6N+2}{N^3(N+1)^3}
       ~,\label{16e}\\ \N\\ 
     \sum_{k=1}^{N-1}\binom{N-1}{k}(-1)^k\frac{S_2(k)}{k}
       &=&-S_{2,1}(N-1)
       ~,\label{7e} \\ \N\\    
     \sum_{k=1}^{N-1}\binom{N-1}{k}(-1)^k\frac{S_2(k)}{k+1}
       &=&-\frac{S_{1,1}(N)}{N}+\frac{S_1(N)}{N^2}
       ~,\label{9e} \\ \N\\    
     \sum_{k=1}^{N-1}\binom{N-1}{k}(-1)^k\frac{S_2(k)}{2+k}
       &=&-\frac{S_{1,1}(N)}{N(N+1)}+\frac{1-N}{N^2}S_{1}(N)+\frac{1}{N+1}
       ~,\label{1e} \\ \N\\    
     \sum_{k=1}^{N-1}\binom{N-1}{k}(-1)^k\frac{S_2(k)}{3+k}
       &=&-\frac{2S_{1,1}(N)}{N(N+1)(N+2)}
        -\frac{N^2+4N-4}{2(N+2)N^2}S_{1}(N)\N\\
       &&+\frac{N+11}{4(N+1)(N+2)}
       ~,\label{2e}\\ \N\\
     \sum_{k=1}^{N-1}\binom{N-1}{k}(-1)^k\frac{S_2(k)}{(1+k)^2}
       &=&\frac{S_{2,1}(N)}{N}-\frac{S_{1,1,1}(N)}{N}
       ~,\label{10e} \\ \N\\
     \sum_{k=1}^{N-1}\binom{N-1}{k}(-1)^k\frac{S_2(k)}{(2+k)^2}
       &=&\frac{S_{2,1}(N)}{N(N+1)}-\frac{S_{1,1,1}(N)}{N(N+1)}
       +\frac{S_{1,1}(N)}{(N+1)^2}-\frac{N+2}{N(N+1)}S_{1}(N) \N\\
       &&+\frac{2N+3}{(N+1)^2}
       ~,\label{3e} \\ \N\\    
     \sum_{k=1}^{N-1}\binom{N-1}{k}(-1)^k\frac{S_2(k)}{(3+k)^2}
       &=&\frac{2S_{2,1}(N)}{N(N+1)(N+2)}-\frac{2S_{1,1,1}(N)}{N(N+1)(N+2)}\N\\
       &&+\frac{(3N+5)S_{1,1}(N)}{(N+1)^2(N+2)^2}
         -\frac{40+38N+9N^2+N^3}{4N(N+1)(N+2)^2}S_{1}(N)\N\\
       &&+\frac{59+66N+18N^2+N^3}{4(N+1)^2(N+2)^2}
       ~,\label{4e}%
\\ \N\\
     \sum_{k=0}^{N-1}\binom{N-1}{k}(-1)^k\frac{S_{1,1}(k+A)}{k+A}
       &=&\frac{B(N,A)}{2}\Biggl[
           \Bigl\{S_1(N+A-1)-S_1(N-1)\Bigr\}^2 \N\\
       &&+S_2(N-1+A)-S_2(N-1)\Bigr]
       ~,\label{12e}\\ \N\\ 
     \sum_{k=0}^{N-1}\binom{N-1}{k}(-1)^k\frac{S_{1,1}(k)}{k}
       &=&\zeta_3-S_3(N-1)
         ~,\label{20e}\\ \N\\ 
     \sum_{k=1}^{N-1}\binom{N-1}{k}(-1)^k\frac{S_3(k)}{k}
       &=&-S_{2,1,1}(N-1)
       ~,\label{8e} \\ \N
\\
     \sum_{k=1}^{N-1}\binom{N-1}{k}(-1)^k\frac{S_3(k)}{k+1}
       &=&-\frac{S_{1,1,1}(N-1)}{N}
       ~,\label{17e} \\ \N
\end{eqnarray}\begin{eqnarray}
     \sum_{k=1}^{N-1}\binom{N-1}{k}(-1)^k\frac{S_3(k)}{2+k}
       &=&-\frac{S_{1,1,1}(N)}{N(N+1)}+\frac{1-N}{N^2}S_{1,1}(N)
       +\frac{S_{1}(N)}{N}\N\\
       &&-\frac{1}{N+1}
       ~,\label{5e} \\ \N\\    
     \sum_{k=1}^{N-1}\binom{N-1}{k}(-1)^k\frac{S_3(k)}{3+k}
       &=&-\frac{2S_{1,1,1}(N)}{N(N+1)(N+2)}
       -\frac{N^2+4N-4}{2(N+2)N^2}S_{1,1}(N)\N\\
       &&+\frac{S_1(N)(N+10)}{4N(N+2)}
         -\frac{N+19}{8(N+1)(N+2)}
       ~,\label{6e} \\ \N\\
     \sum_{k=0}^{N-1}\binom{N-1}{k}(-1)^k\frac{S_{1,1,1}(k+A)}{k+A}
       &=&-\frac{1}{6}\frac{d^3}{dN^3}B(N,A)
       ~,\label{21e}\\ \N\\ 
     \sum_{k=0}^{N-1}\binom{N-1}{k}(-1)^k\frac{S_{1,1,1}(k)}{k}
       &=&\frac{2}{5}\zeta_2^2-S_4(N-1)
           ~,\label{22e}\\ \N\\ 
     \sum_{k=0}^{N-1}\binom{N-1}{k}(-1)^k\frac{S_{2,1}(k)}{k}
       &=&\frac{7}{10}\zeta_2^2-S_{2,2}(N-1)
          ~,\label{23e} 
    \end{eqnarray}
\vspace{-5mm}
    \begin{eqnarray}
   && \sum_{k=0}^{N-1}\binom{N-1}{k}(-1)^k \Biggl\{
       -\frac{\zeta_3}{k}+2\frac{\zeta_2}{k^2}
       -2\frac{S_1(k)}{k^3}-\frac{S_{1,1}(k)}{k^2}
       \Biggr\} \N\\
    &=&-\frac{7}{10}\zeta_2^2
     +\zeta_3S_1(N-1)-2\zeta_2S_{1,1}(N-1)
       +S_{1,3}(N-1)+2S_{1,1,2}(N-1)
       ~.\label{l1}
   \end{eqnarray}
  \subsection{\bf\boldmath Sums involving Binomials with Two Free Indexes}
   \begin{eqnarray}
     \sum_{k=0}^{L+1}\binom{L+1}{k}\frac{(-1)^k}{N-L+k}
       &=&B(N-L,L+2)
       ~,\label{f1}\\ \N\\ 
     \sum_{k=0}^{L+1}\binom{L+1}{k}\frac{(-1)^k}{(N-L+k)^2}
       &=&B(N-L,L+2)\Bigl[S_1(N+L) \N\\
        &&-S_1(N-L-1)\Bigr]
       ~,\label{f2}\\ \N\\ 
     \sum_{k=0}^{L+1}\binom{L+1}{k}(-1)^k\frac{S_1(N-L+k)}{N-L+k}
       &=&B(N-L,L+2)\Bigl[S_1(N+1)-S_1(L+1)\Bigr]
       ~,\label{g1} \\    
     \sum_{k=0}^{L+1}\binom{L+1}{k}(-1)^k\frac{S_1(N-L+k)}{1+N-L+k}
       &=&B(N-L+1,L+2)\Bigl[S_1(N-L)-S_1(L+1)\Bigr] \N\\
       &=&-\frac{d}{dL}B(N-L+1,L+2)
       ~,\label{g2}
\end{eqnarray}\begin{eqnarray}
     \sum_{k=1}^{\infty}\frac{B(k+\ep/2,N+1)}{N+k}
       &=&(-1)^N\Bigl[2S_{-2}(N)+\zeta_2\Bigr] \N\\
       &+&\frac{\ep}{2}(-1)^N\Bigl[
          -\zeta_3+\zeta_2S_1(N)+2S_{1,-2}(N)-2S_{-2,1}(N)
          \Bigr] \N\\
       &+&\frac{\ep^2}{4}(-1)^N\Bigl[
           \frac{2}{5}\zeta_2^2-\zeta_3S_1(N)+\zeta_2S_{1,1}(N)
          +2\Bigl\{S_{1,1,-2}(N)\N\\
        &&+S_{-2,1,1}(N)
          -S_{1,-2,1}(N)\Bigr\}
          \Bigr]+O(\ep^3) ~,\label{j1}
   \end{eqnarray}

\vspace{5mm}\noindent
{\bf Acknowledgments.}~~We would like to thank W.L.~van Neerven and J.~Smith 
for the possibility to compare results on the level of diagrams and useful 
discussions. We thank P. Paule and C. Schneider for their interest in the
present work. This work was supported in part by DFG Sonderforschungsbereich 
Transregio 9, Computergest\"utzte Theoretische
Teilchenphysik.


\newpage
\section{Figures}
\label{sec:fig}

\vspace{2cm}
\newpage
\newpage
\end{document}